\newif\ifonecolumn
\onecolumnfalse
\ifonecolumn
  \documentclass[journal,draftclsnofoot,onecolumn, 12pt]{IEEEtran}
\else
  \documentclass[journal]{IEEEtran}
\fi

\usepackage[T1]{fontenc}

\usepackage{amsmath}
\usepackage{cite}
\usepackage{array}
\usepackage{stfloats}
\usepackage{url}
\usepackage{amsmath,amssymb}
\usepackage{epsfig,fancyhdr}
\usepackage{CJKutf8}
\usepackage{times}
\usepackage{graphicx}
\usepackage{caption}
\usepackage{subcaption}
\usepackage{multirow}
\usepackage{stfloats}
\usepackage{color}
\usepackage[dvipsnames]{xcolor}

\usepackage{arydshln} 

\usepackage{bm}

\newtheorem{thm}{Theorem}

\newtheorem{defin}{Definition}
\newtheorem{lemma}{Lemma}

\newtheorem{eg}{Example}
\newtheorem{remark}{Remark}
\newcommand{\Mod}[1]{\text{mod}\, #1}

\begin{document}

\title{Enhanced Cross Z-Complementary Set and Its Application in Generalized Spatial Modulation}

\author{Zhen-Ming Huang,~\IEEEmembership{Graduate Student Member,~IEEE}, 
Cheng-Yu~Pai,~\IEEEmembership{Member,~IEEE}, 
Zilong Liu,~\IEEEmembership{Senior Member,~IEEE},
and Chao-Yu~Chen,~\IEEEmembership{Senior Member,~IEEE}
\thanks{Z.-M. Huang is with the Institute of Computer and Communication Engineering and the Department of Engineering Science,  National Cheng Kung University, Tainan 701, Taiwan, R.O.C. (e-mail: n98101012@gs.ncku.edu.tw).

  C.-Y. Pai is with the Department of Engineering Science, National Cheng Kung University, Tainan 701, Taiwan, R.O.C. (e-mail: n98081505@gs.ncku.edu.tw).
  
  Zilong Liu is with the School of Computer Science and Electronic Engineering, University of Essex, UK. (e-mail: zilong.liu@essex.ac.uk).
  
  C.-Y. Chen is with the Department of Electrical Engineering and the Institute of Computer and Communication Engineering, National Cheng Kung University, Tainan 701, Taiwan, R.O.C. (e-mail: super@mail.ncku.edu.tw).
  }}

\maketitle

\begin{abstract}
Generalized spatial modulation (GSM) is a novel multiple-antenna technique offering flexibility among spectral efficiency, energy efficiency, and the cost of RF chains. In this paper, a novel class of sequence sets, called enhanced cross Z-complementary set (E-CZCS), is proposed for efficient training sequence design in broadband GSM systems. Specifically, an \mbox{E-CZCS} consists of multiple CZCSs possessing front-end and tail-end zero-correlation zones (ZCZs), whereby any two distinct CZCSs have a tail-end ZCZ when a novel type of cross-channel aperiodic correlation sums is considered. The theoretical upper bound on the ZCZ width is first derived, upon which optimal E-CZCSs with flexible parameters are constructed. For optimal channel estimation over frequency-selective channels, we introduce and evaluate a novel GSM training framework employing the proposed \mbox{E-CZCSs}. 
\end{abstract}


\section{Introduction}
\IEEEPARstart{G}{olay} complementary pair (GCP), found by Marcel J. E. Golay in the middle of the 20th century, is characterized by the property that the aperiodic autocorrelation sum of the two constituent sequences is zero at every non-zero time-shift \cite{Golay}. In 1972, Tseng and Liu extended the concept of GCP to Golay complementary set (GCS), each consisting of more than two constituent sequences \cite{Golay_sets}. Additionally, a collection of GCSs was introduced, called the mutually orthogonal complementary set (MOCS), where any two distinct GCSs in an MOCS have zero aperiodic cross-correlation sums for all time-shifts. In \cite{N_shift}, complete complementary code (CCC) was introduced as optimal MOCS with the maximum set size. Owing to these ideal autocorrelation and cross-correlation properties, complementary pairs/sets of sequences have been employed in many communications applications including synchronization \cite{CS_sync}, channel estimation \cite{CS_CE, Wang07}, interference suppression \cite{Bell_CDMA,Chen_CDMA,Chen2007book,Liu2014,Liu2015}, peak-to-mean power control \cite{Nee2, Golay_RM, Paterson_00, Super_16, Super_18}, cell search \cite{ChenICC08}, and MIMO radar \cite{Li2010,Tang2014}. As a generalization of MOCSs and CCCs, Z-complementary code set (ZCCS) having zero correlation zone (ZCZ) was proposed in \cite{ZCP-1st}.

On the other hand, spatial modulation (SM) has received tremendous research attention in recent years as a novel multiple-antenna technique. In SM, there is only one radio-frequency (RF) chain, whereby one transmit antenna (TA) is activated at each time-slot \cite{SM_1,SM_2,SM_3,SM_4,SM_5,SM_6,SM_7}. Because of this, SM enjoys zero inter-antenna interference (IAI), lower energy consumption, and reduced transceiver complexity. For a long time, efficient channel estimation schemes for SM systems in frequency selective channels were missing. In 2020, \mbox{Liu {\it et al.}} proposed cross Z-complementary pair (CZCP) for optimal sparse training matrix design in broadband SM systems \cite{CZCP-1st}. Several constructions of CZCPs with larger ZCZ widths and more flexible lengths have been proposed in \cite{CZCP-1st, Fan_20, Adhikary_20, Huang_20, Yang_21, Zeng_22, Das_22_ISIT, Zhang_22}. However, the ZCZ width of every CZCP is theoretically upper bounded by a half of its sequence length. In \cite{Huang_22, Huang_22_ISIT}, cross Z-complementary set (CZCS) which can tolerate larger delay spreads was developed.

Unlike SM, generalized spatial modulation (GSM) system has been proposed for a higher spectral efficiency as it allows two or more active TAs at the same time \cite{GSM_first, GSM_1, GSM_2, GSM_3, GSM_4}. To be specific, the transmitter of a GSM system is equipped with a few RF chains less than the number of TAs. During each transmission, a GSM symbol is modulated using two information parts. The message bits of the first information part are used to select the antenna activation patterns, whereas the second part carries message bits for selecting specific constellation points over those activated TAs. Therefore, GSM provides an excellent trade-off between the spectral efficiency and the cost of RF chains, while retaining most of the advantages of SM.

Training sequence design for GSM is a more challenging task. First,  dense training sequences designed for the traditional multiple-input multiple-output (MIMO) in \cite{MSE_bound, dense_training_2, dense_training_3} cannot be used since only a few GSM TAs are activated at each time-slot. Recently, symmetrical Z-complementary code set (SZCCS) was proposed in \cite{Zhou_23} for GSM training design. It is noted that SZCCS is a subclass of ZCCSs with symmetric ZCZ properties for its autocorrelation and cross-correlation sums. However, the proposed GSM training framework in \cite{Zhou_23} has an additional overhead for IAI mitigation incurred by zero-padding. Consequently, their approach suffers from a reduced training efficiency. Motivated by \cite{CZCP-1st, Huang_22}, we aim to go beyond the CZCP and CZCS by introducing new sequence properties for more efficient training design in GSM.

In this paper, we propose a novel family of CZCSs, called enhanced cross Z-complementary set (E-CZCS), each consisting of multiple CZCSs and any two distinct CZCSs have a tail-end ZCZ when a special type of cross-channel aperiodic correlation sums is considered. More specifically, such a tail-end ZCZ is required for the cross-correlation sums between the $n$-th constituent sequence  of one CZCS and the $(n+1)_{\Mod N}$-th constituent sequence of the other, where $N$ refers to the total number of constituent sequences. The major contributions of this paper are summarized as follows:

\begin{itemize}
  \item We extend the concept of CZCS to E-CZCS by incorporating the aforementioned cross-channel aperiodic ZCZ property. Additionally, we derive an upper bound on the width of the ZCZ which allows us to define optimal E-CZCS with maximum ZCZ width.
  \item Two constructions of E-CZCSs are proposed. The first construction is based on MOCSs, CCCs, and ZCCSs. The second construction is based on generalized Boolean functions. Both constructions can generate optimal binary E-CZCSs with various set sizes.
  \item We present a novel training framework employing the proposed E-CZCSs for broadband GSM systems. The proposed GSM training framework can achieve optimal channel estimation over frequency-selective channels. Both IAI and ISI can be eliminated, thanks to the unique correlation properties of the proposed E-CZCSs.
      
  \item Simulations show that the proposed E-CZCS-based training scheme can achieve the minimum channel estimation mean square error (MSE) and outperform other classes of sequences, such as SZCCSs, ZCCSs, and Zadoff-Chu sequences.

  \end{itemize}

The rest of this paper is organized as follows. In Section \ref{sec:def}, we first introduce some necessary notations,  definitions, and the GSM system. In Section \ref{sec:GCZCS}, we deﬁne the E-CZCS and its correlation properties, and then propose two constructions of E-CZCSs. Section \ref{sec:Proposed Training Framework} describes the requirements for training design in the GSM system and proposes a novel training framework based on E-CZCSs. The performance comparison is provided in Section \ref{sec:Numerical Evaluation}. Finally, concluding remarks are drawn in Section \ref{sec:conclusion}.

\section{Preliminaries and Definitions}\label{sec:def}
First, we introduce some notations which are used throughout this paper.
\subsection{Notations}
\begin{itemize}
\item ``$\bm a\|\bm b$'' denotes the concatenation of sequences $\bm a$ and $\bm b$;
\item ``$+$'' and ``$-$'' denote $1$ and $-1$, respectively;
\item $\xi_q=e^{2\pi j/q}$ is a primitive complex $q$th root of unity;
\item $\bm X^{*}$ denotes the complex conjugate of the matrix $\bm X$;
\item $\bm X^T$ denotes the transpose of the matrix $\bm X$;
\item $\bm X^{H}$ denotes the Hermitian of the matrix $\bm X$;
\item $\lfloor \cdot \rfloor$ denotes the floor operation;
\item $(\cdot)_{\Mod N}$ denotes the modulo operation with respect to a positive integer $N$;
\item $\text{Tr}(\bm X)$ denotes the trace of the square matrix $\bm X$;
\item $\bm{X}^{(L)}$ is the matrix where each row is the cyclic-shift ($L$ elements to the right) of the corresponding row in $\bm{X}$.
\end{itemize}

Let ${\bm{s}_{0}}=(s_{0,0},s_{0,1},\ldots,s_{0,L-1})$ and ${\bm{s}_{1}}=(s_{1,0},s_{1,1},\ldots,s_{1,L-1})$ denote two complex-valued sequences of length $L$. For any integer displacement $u$, the {\em aperiodic cross-correlation function} (ACCF) of ${\bm{s}_{0}}$ and ${\bm{s}_{1}}$ is defined as
\begin{equation}
\begin{aligned}
  \rho({\bm{s}_{0}},{\bm{s}_{1}};u)=\begin{cases}
\sum\limits_{k=0}^{L-1-u}{s_{0,k+u}s_{1,k}^{*}},\quad 0 \leq u\leq L-1;\\
\sum\limits_{k=0}^{L-1+u}{s_{0,k}s_{1,k-u}^{*}},\quad -L+1 \leq u<0.
\end{cases}
\end{aligned}
  \label{eq:across}
\end{equation}
When $\bm{s}_{0}= \bm{s}_{1}$, the function $\rho({\bm{s}_{0}},{\bm{s}_{1}};u)=\rho(\bm{s}_{0}; u)$ is referred to as the {\em aperiodic autocorrelation function} (AACF) of $\bm{s}_{0}$. For periodic correlations, the {\em periodic cross-correlation function} (PCCF) of ${\bm{s}_{0}}$ and ${\bm{s}_{1}}$ at time-shift $u$ is defined as
\begin{equation}
\begin{aligned}
  \phi({\bm{s}_{0}},{\bm{s}_{1}};u)=\begin{cases}
\sum\limits_{k=0}^{L-1}{s_{0,(k+u)_{\Mod L}}s_{1,k}^{*}},\quad 0 \leq u\leq L-1;\\
\sum\limits_{k=0}^{L-1}{s_{0,k}s_{1,(k-u)_{\Mod L}}^{*}},\quad -L+1 \leq u<0.
\end{cases}
\end{aligned}
  \label{eq:pcross}
\end{equation}
Accordingly, the {\em periodic autocorrelation function} (PACF) of $\bm{s}_{0}$ is denoted by $\phi(\bm{s}_{0}, \bm{s}_{0}; u)=\phi(\bm{s}_{0}; u)$. Then, the AACFs and PCCFs are related by
\begin{equation}
\begin{aligned}
  &\phi({\bm{s}_{0}},{\bm{s}_{1}};u)
  &=\begin{cases}
\rho({\bm{s}_{0}},{\bm{s}_{1}};u),\hspace*{4em} u=0;\\
\rho({\bm{s}_{0}},{\bm{s}_{1}};u)+\rho^{*}({\bm{s}_{1}},{\bm{s}_{0}};L-u),\\
\hspace*{8em}\quad 0 < u\leq L-1;\\
\rho^{*}({\bm{s}_{1}},{\bm{s}_{0}};-u)+\rho({\bm{s}_{0}},{\bm{s}_{1}};L+u),\\
\hspace*{8em}\quad -L+1 \leq u<0.
\end{cases}
\end{aligned}
  \label{eq:recross}
\end{equation}

\begin{defin}
For a set of $N$ complex sequences $\mathcal{S} = \{\bm{s}_{0}, \bm{s}_{1}, . . . , \bm{s}_{N-1}\}$ with length $L$, if
\begin{equation}
\begin{aligned}
\phi(\bm{s}_{i},\bm{s}_{j};u)=
\begin{cases}
     0,  & 1\leq|u|\leq Z, 0 \leq i=j \leq N-1;\\
     0,  & |u|\leq Z, 0 \leq i\neq j \leq N-1,
 \end{cases}
 \label{eq:ZCZ_seq}
\end{aligned}
\end{equation}
then the set $\mathcal{S}$ is called an $(N,L,Z)$-ZCZ sequence set where $Z$ is referred to as the width of ZCZ. The following lemma shows an upper bound among the parameters of the ZCZ sequence set.
\end{defin}
\begin{lemma}\cite{Tang00}\label{eq:ZCZ_bound}
For an $(N,L,Z)$-ZCZ sequence set, there is a well-known theoretical upper bound, called {\em Tang-Fan-Matsufuji bound}, given as $Z \leq {L}/{N}-1$. A ZCZ sequence set is said to be optimal if the Tang-Fan-Matsufuji bound with equality is achieved. However, for binary case, the upper bound on ZCZ width is conjectured to be $Z \leq {L}/{(2N)}$.
\end{lemma}

Consider a set of $M$ sequence sets $\mathcal{S}=\{S^m|0\leq m \leq M-1\}$ where each constitute set $S^m=\{\bm{s}_{n}^{m}|0\leq n \leq N-1\}$ consists of $N$ sequences of length $L$. For $S^{m_{1}},S^{m_{2}}\in \mathcal{S}$ and $0\leq m_{1}, m_{2}\leq M-1$, we define
\begin{equation}
\begin{aligned}
  \rho(S^{m_{1}},S^{m_{2}};u)\triangleq\sum_{n=0}^{N-1}\rho({\bm{s}_{n}^{m_{1}}},{\bm{s}_{n}^{m_{2}}};u).
\end{aligned}
  \label{eq:ACCF sums}
\end{equation}

\begin{defin}
A set of $M$ sequence sets $\mathcal{S}=\{S^m|0\leq m \leq M-1\}$ is addressed as {\em Z-complementary code set}, denoted by $(M,N,L,Z)$-ZCCS, if
\begin{equation}
\begin{aligned}
 \rho(S^{m_1},S^{m_2};u)&=\sum_{n=0}^{N-1}\rho(\bm s_{n}^{m_1},\bm s_{n}^{m_2};u)\\
 &=
 \begin{cases}
     NL, & u=0, m_1=m_2; \\
     0,  & 0<|u|<Z, m_1=m_2;\\
     0,  & |u|<Z, m_1\neq m_2
 \end{cases}
 \label{eq:ZCCS}
\end{aligned}
\end{equation}
where $M$ is the set size, $N$ is the number of sequences in each $S^m$, $L$ is the sequence length, and $Z$ is the width of ZCZ. For an $(M,N,L,Z)$-ZCCS, a theoretical upper bound on the set size is given as $M\leq N\lfloor {L}/{Z}\rfloor$. An $(M,N,L,Z)$-ZCCS is called optimal if the equality is achieved.

When $Z=L$, the set $\mathcal{S}$ is referred to as a {\em mutually orthogonal complementary set}, denoted by $(M,N,L)$-MOCS. Specifically, each constituent sequence set $S^m$ reduces to a GCS and any two GCSs in set $\mathcal{S}$ are mutually orthogonal. Likewise, the upper bound on the set size for an MOCS satisﬁes the inequality $M \leq N\lfloor L/L\rfloor =N$. If $M = N$, the set $\mathcal{S}$ is called a {\em complete complementary code}, denoted by $(M,L)$-CCC.

\begin{defin}\cite{Zhou_23}
A set of $M$ sequence sets $\mathcal{S}=\{S^m|0\leq m \leq M-1\}$ is called a {\em symmetrical Z-complementary code set}, denoted by $(M,N,L,Z)$-SZCCS, if
\begin{equation}
\begin{aligned}
\rho(S^{m_1},S^{m_2};u)=
\begin{cases}
\sum\limits_{n=0}^{N-1}\rho(\bm s_{n}^{m};u)=0,\\
\hspace*{3em}  \text{for\,}|u|\in\mathcal {T}_1 \cup \mathcal {T}_2, m_1=m_2=m;\\
\sum\limits_{n=0}^{N-1}\rho(\bm s_{n}^{m_1},\bm s_{n}^{m_2};u)=0, \\
\hspace*{3em}  \text{for\,}|u|\in \mathcal {T}_1 \cup \mathcal {T}_2\cup \{0\}, m_1\neq m_2
\end{cases}
 \label{eq:SZCCS}
\end{aligned}
\end{equation}
where $\mathcal{T}_1\triangleq\{1,2,\ldots,Z\}$ and $\mathcal{T}_2\triangleq\{L-Z,L-Z+1,\ldots,$ $L-1\}$.
\end{defin}

\subsection{Generalized Boolean Functions}\label{sec:GBF}

Let $f$ be a function of $m$ $\mathbb{Z}_2$-valued variables $x_1, x_2, \ldots, x_m$ mapping from $\mathbb{Z}_2^m$ to $\mathbb{Z}_q$, denoted by
\begin{equation}
f:(x_1,x_2,\ldots, x_{m})\in\mathbb{Z}_2^m \rightarrow f(x_1,x_2,\ldots, x_{m})\in\mathbb{Z}_q.
 \label{eq:GBF1}
\end{equation}
The function $f$ is addressed as the generalized Boolean function. For a $q$-ary generalized Boolean function $f$, the associated sequence $\bm f \in \mathbb{Z}_{q}^{2^m}$ is given by
\begin{equation}
\bm f = (f_0, f_1,\ldots, f_{2^{m}-1})
 \label{eq:GBF2}
\end{equation}
where $f_i=f(i_1,i_2,\ldots,i_m)$ and $(i_1,i_2,\ldots,i_m)$ is the binary representation of the integer $i=\sum_{k=1}^{m}i_{k}2^{k-1}$. Note that
 $i_1$ is the least significant bit.

Also, we define the complex-valued sequence associated with a generalized Boolean function $f$ to be
\begin{equation}
\zeta_{q}(\bm{f})\triangleq(\xi_{q}^{f_0},\xi_{q}^{f_1},\ldots,\xi_{q}^{f_{2^{m}-1}}).
 \label{eq:GBF3}
\end{equation}
Let $m = 4$ and $q=4$ as an example. The sequence $\bm f$ associated with $f=2x_2+x_{1}x_{2}$ is
\begin{equation*}
\begin{aligned}
\bm f =2\bm{x}_2+\bm{x}_{1}\bm{x}_{2}&= (f(0,0,0,0),f(1,0,0,0),\ldots,f(1,1,1,1))\\
&=(0,0,2,3,0,0,2,3,0,0,2,3,0,0,2,3)
\end{aligned}
 \label{eq:GBF4}
\end{equation*}
and the complex modulated sequence 
\begin{align*}
\zeta_{4}(\bm{f})&=(\xi_{4}^{f_{0}},\xi_{4}^{f_{1}},\ldots,\xi_{4}^{f_{15}})\\
&=(1,1,-1,-j,1,1,-1,-j,1,1,-1,-j,1,1,-1,-j).
\end{align*}

\subsection{GSM}

Consider a single-carrier GSM (SC-GSM) system over frequency-selective channels as depicted in Fig.~\ref{GSM_fig}. We denote $N_t$, $N_{r}$, and $N_{a}$ as the number of TAs, receive antennas (RAs), and RF chains, respectively. An $N_{a}\times N_t$ switch is needed to connect the RF chains to the TAs. During each time-slot $k$, $N_{a}$ of the $N_t$ TAs are activated and the corresponding constellation symbols from QAM/PSK modulation $\mathcal{M}$ are transmitted on the activated TAs, while the remaining $N_{t}-N_{a}$ antennas are kept inactive. Specifically, $\lfloor \log_{2}\binom{N_{t}}{N_{a}}\rfloor$ information bits, denoted by $\bm{p}_{k}$, are used for selecting $N_{a}$ antennas based on activation pattern mapping. Additionally, we denote an $N_{t}\times 1$ vector $\bm{s}=(0\cdots 1 \cdots 0 \cdots 1)^{T}$ as an antenna activation pattern where $1$'s correspond to the active antennas and $0$'s correspond to the silent antennas. On the other hand, $N_{a}\lfloor \log_{2} |\mathcal{M}| \rfloor$ bits, denoted by $\bm{q}_{k}$, are mapped into a constellation from alphabet $\mathcal{M}$ through $N_{a}$ active antennas. Therefore, the number of bits conveyed per symbol period is given by $\lfloor \log_{2}\binom{N_{t}}{N_{a}}\rfloor + N_{a}\lfloor \log_{2} |\mathcal{M}| \rfloor$.

\begin{figure*}[!t]
\centering
\begin{center}
\extrarowheight=3pt
\includegraphics[width=130mm]{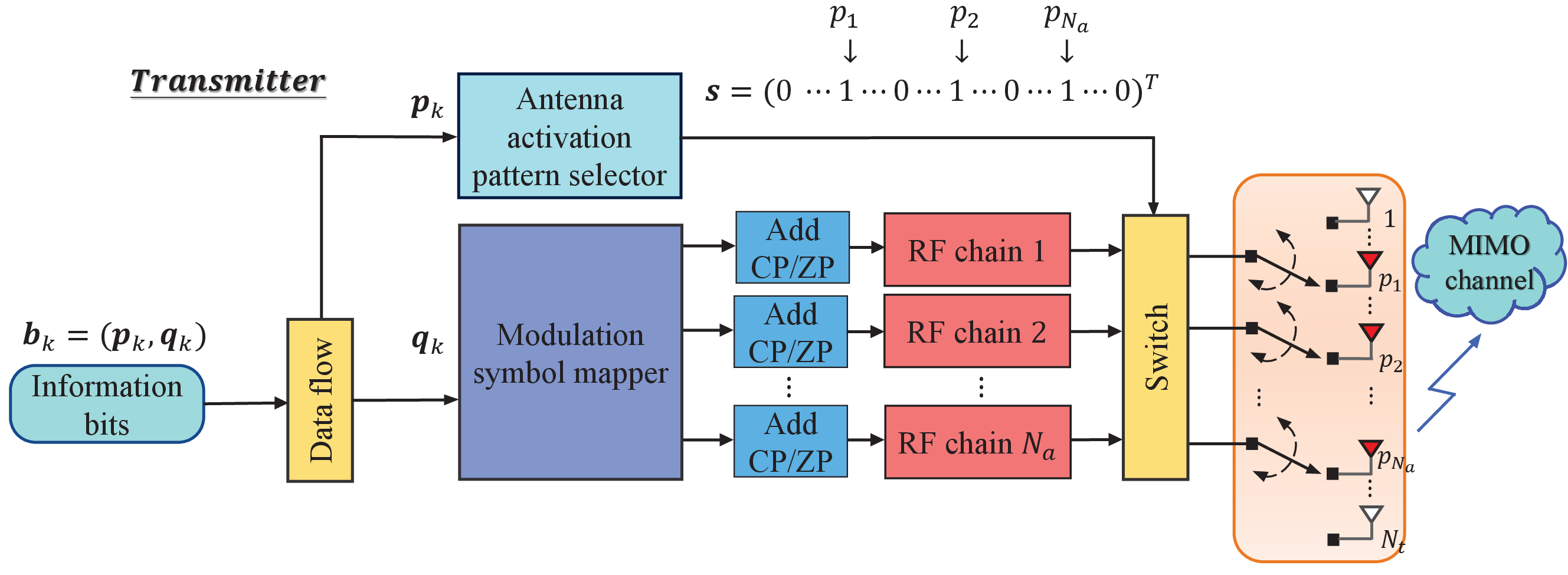}
\caption{A generic transmitter structure of the SC-GSM system.\label{GSM_fig}}
\end{center}
\end{figure*}

\begin{eg}
Let us consider an SC-GSM system with $N_{t}=4$ and $N_{a}=2$ using BPSK modulation (i.e., $|\mathcal{M}|=2$). The $\binom{4}{2}=6$ possible activation patterns are shown as follows:
$(1,1,0,0)^{T}$, $(1,0,1,0)^{T}$, $(1,0,0,1)^{T}$, $(0,1,1,0)^{T}$, $(0,1,0,1)^{T}$, $(0,0,1,1)^{T}$. However, only $4$ activation patterns are selected since $\lfloor \log_{2}\binom{4}{2}\rfloor=2$. Then, the set of chosen activation patterns in this example is given by $\{(1,1,0,0)^{T},(0,1,1,0)^{T},(1,0,0,1)^{T},(0,0,1,1)^{T}\}$. Table~\ref{table1} shows a mapping from $2$ information bits to the set of chosen activation patterns.  Assume that the message bits $\left(0101001110101111\right)$ are sent. There are $4$ GSM symbols of which each symbol consists of $\lfloor \log_{2}\binom{4}{2}\rfloor + 2\lfloor \log_{2} |2| \rfloor = 4$ bits. We have $\bm b_1=\left(0101\right)$, $\bm b_2=\left(0011\right)$, $\bm b_3=\left(1010\right)$, and $\bm b_4=\left(1111\right)$. Taking the first symbol for example, we have $\bm b_1=\left(\bm p_1, \bm q_1\right)$ where $\bm p_1=\left(01\right)$ and $\bm q_1=\left(01\right)$. It indicates that TA $2$ and TA $3$ are activated to transmit the BPSK symbol “$1$” and “$-1$”, respectively. Therefore, the first GSM symbol can be expressed as $\left(0,+,-,0\right)^{T}$. Then, the SC-GSM block for the $4$ GSM symbols can be formulated as follows:
\begin{align}
\left(
	\begin{array}{c|c|c|c}
	0 & - & - & 0  \\
    + & - & 0 & 0  \\
    - & 0 & 0 & -  \\
    0 & 0 & + & -
	\end{array}
\right).
\end{align}
\end{eg}

\begin{table}[t]
\begin{scriptsize}
\centering
\extrarowheight=3pt
\caption{An Example of Antenna Activation Patterns with $N_{t}=4$ and $N_{a}=2$  \label{table1}}
\begin{center}
\begin{tabular}{|c|c|c|}
\hline
\multicolumn{3}{|c|}{GSM Mapping Rule}                           \\ \hline
Information bits   & Antenna activation patterns     & Note           \\ \hline
00 & $(1,1,0,0)^{T}$ & \begin{tabular}[c]{@{}c@{}}Active antennas : 1, 2; \\ Inactive antennas : 3, 4\end{tabular}  \\ \hline
01 & $(0,1,1,0)^{T}$ & \begin{tabular}[c]{@{}c@{}}Active antennas : 2, 3; \\ Inactive antennas : 1, 4\end{tabular}  \\ \hline
10 & $(1,0,0,1)^{T}$ & \begin{tabular}[c]{@{}c@{}}Active antennas : 1, 4; \\ Inactive antennas : 2, 3\end{tabular}  \\ \hline
11 & $(0,0,1,1)^{T}$ & \begin{tabular}[c]{@{}c@{}}Active antennas : 3, 4; \\ Inactive antennas : 1, 2\end{tabular}  \\ \hline
\end{tabular}
\end{center}
\end{scriptsize}
 \end{table}

\section{Enhanced Cross Z-Complementary Sets}\label{sec:GCZCS}

In this section, we will provide the definition and the optimality of the E-CZCS and then demonstrate two novel constructions.

Consider a set of $M$ sequence sets $\mathcal{G}=\{G^m|0\leq m \leq M-1\}$ where each constitute set $G^m=\{\bm{g}_{n}^{m}|0\leq n \leq N-1\}$ is composed of $N$ sequences of length $L$. For $G^{m_{1}},G^{m_{2}}\in \mathcal{G}$ with $0\leq m_{1}, m_{2}\leq M-1$, we define a special type of aperiodic cross-correlation sum as follows:
\begin{equation}
\begin{aligned}
  \hat{\rho}(G^{m_{1}},G^{m_{2}};u)\triangleq\sum_{n=0}^{N-1}\rho(\bm g_{n}^{m_{1}},\bm g_{(n+1)_{\Mod N}}^{m_{2}};u).
\end{aligned}
  \label{eq:adjacent ACCF sums}
\end{equation}
\end{defin}
Note that (\ref{eq:adjacent ACCF sums}) is different from the cross-correlation sum defined in (\ref{eq:ACCF sums}). Taking $M=2$ and $N=4$ for example, we assume $\mathcal{G}=\{G^0, G^1\}$ where $G^0=\{\bm{g}_{0}^{0},\bm{g}_{1}^{0},\bm{g}_{2}^{0},\bm{g}_{3}^{0}\}$ and $G^1=\{\bm{g}_{0}^{1},\bm{g}_{1}^{1},\bm{g}_{2}^{1},\bm{g}_{3}^{1}\}$. Then, we have
\begin{equation*}
\begin{aligned}
  &\hat{\rho}(G^{0},G^{0};u)=\sum_{n=0}^{3}\rho(\bm g_{n}^{0},\bm g_{(n+1)_{\Mod 4}}^{0};u)\\&=\rho(\bm g_{0}^{0},\bm g_{1}^{0};u)+\rho(\bm g_{1}^{0},\bm g_{2}^{0};u)+\rho(\bm g_{2}^{0},\bm g_{3}^{0};u)+\rho(\bm g_{3}^{0},\bm g_{0}^{0};u)
\end{aligned}
\end{equation*}
and
\begin{equation*}
\begin{aligned}
  &\hat{\rho}(G^{0},G^{1};u)=\sum_{n=0}^{3}\rho(\bm g_{n}^{0},\bm g_{(n+1)_{\Mod 4}}^{1};u)\\&=\rho(\bm g_{0}^{0},\bm g_{1}^{1};u)+\rho(\bm g_{1}^{0},\bm g_{2}^{1};u)+\rho(\bm g_{2}^{0},\bm g_{3}^{1};u)+\rho(\bm g_{3}^{0},\bm g_{0}^{1};u).
\end{aligned}
\end{equation*}

Then we can define the E-CZCS based on the cross-correlation sums given in (\ref{eq:ACCF sums}) and (\ref{eq:adjacent ACCF sums}).

\begin{defin}[Enhanced Cross Z-Complementary Set]\label{GCZCS_defin_og}
For positive integers $M$, $N$, $L$, and $Z$ with $Z \leq L$, we denote $\mathcal{T}_1\triangleq\{1,2,\ldots,Z\}$ and \mbox{$\mathcal{T}_2\triangleq\{L-Z,L-Z+1,\ldots,L-1\}$} as two distinct intervals. Let $\mathcal{G}=\{{G}^{m}| 0\leq m \leq M-1\}$ be a set of $M$ sequence sets and $G^m=\{\bm{g}_{n}^{m}|0\leq n \leq N-1\}$ where $\bm{g}_{n}^{m}$ is a sequence of length $L$. Then, the set $\mathcal{G}$ is addressed as a {\it Enhanced cross Z-complementary set}, denoted by $(M,N,L,Z)$-E-CZCS, if it satisfies the following two conditions:
\begin{equation}
\begin{aligned}
\text{(C1):\hspace*{1em}}&\rho(G^{m_{1}},G^{m_{2}};u)=\sum_{n=0}^{N-1}\rho(\bm g_{n}^{m_1},\bm g_{n}^{m_2}; u)\\&=
\begin{cases}
0,  & \text{for all\,}|u|\in\left(\mathcal {T}_1 \cup \mathcal {T}_2\right)\cap \mathcal {T}\footnotemark, \\
&\hspace*{2em}0\leq m_1=m_2 \leq M-1; \\
0,   & \text{for all\,}|u|\in\mathcal {T}_1 \cup \mathcal {T}_2 \cup \{0\}, \\
&\hspace*{2em}0\leq m_1\neq m_2 \leq M-1;
\end{cases}\\
\text{(C2):\hspace*{1em}}&\hat{\rho}(G^{m_{1}},G^{m_{2}};u)=\sum_{n=0}^{N-1}\rho(\bm g_{n}^{m_1},\bm g_{(n+1)_{\Mod N}}^{m_2};u)=0, \\ &\text{for all\,} |u|\in \mathcal{T}_2 \,\,\text{and any\,}m_1,m_2\in\{0,1,\ldots,M-1\}
\end{aligned}\label{GCZCS_defin}
\footnotetext{If $Z$ equals to $L$, we have $\mathcal{T}_1=\{1,2,\ldots,L\}$ and $\mathcal {T}_2=\{0,1,\ldots,L-1\}$. Consequently, $\mathcal{T}_1 \cup \mathcal{T}_2=\{0,1,\ldots,L\}$. However, we have $\rho(G^{m_{1}},G^{m_{2}};0)=NL$ for $0\leq m_1=m_2 \leq M-1$. So we have to exclude $u\neq0$. Therefore, an extra condition of the intersection with  $\mathcal{T}=\{1,2,\ldots,L-1\}$ is needed.}
\end{equation}
where $\mathcal{T}=\{1,2,\ldots,L-1\}$.

If $M=1$, i.e., $m_1=m_2=0$, then the E-CZCS reduces to the {\it cross Z-complementary set}, denoted by $(N,L,Z)$-CZCS. It also implies each $G^{m_1}$ in an E-CZCS is a CZCS. (C1) means the correlation sum $\rho(G^{m_{1}},G^{m_{2}};u)$ has symmetric ZCZs over $\mathcal{T}_1$ and $\mathcal{T}_2$. And (C2) indicates that the cross-correlation sum $\hat{\rho}(G^{m_{1}},G^{m_{2}};u)$ has a tail-end ZCZ for shifts over $\mathcal{T}_2$. Besides, from condition (C1), the E-CZCS is also a ZCCS and a SZCCS. However, the ZCCS and SZCCS do not take the condition (C2) into account. Therefore, the E-CZCS can include CZCS, ZCCS, and SZCCS as special cases. 

\begin{remark}\label{GCZCS2CCC}
For an $(M,N,L,Z)$-E-CZCS with $Z\geq L/2$, we have $\mathcal{T}_1\cup \mathcal{T}_2=\{1,2,\ldots,L-1\}$. Therefore, (C1) in (\ref{GCZCS_defin}) implies that an $(M,N,L,Z)$-E-CZCS is also an \mbox{$(M,N,L)$-MOCS}.
\end{remark}
\end{defin}

\begin{figure*}[!t]
\centering
\begin{center}
\extrarowheight=3pt
\includegraphics[width=160mm]{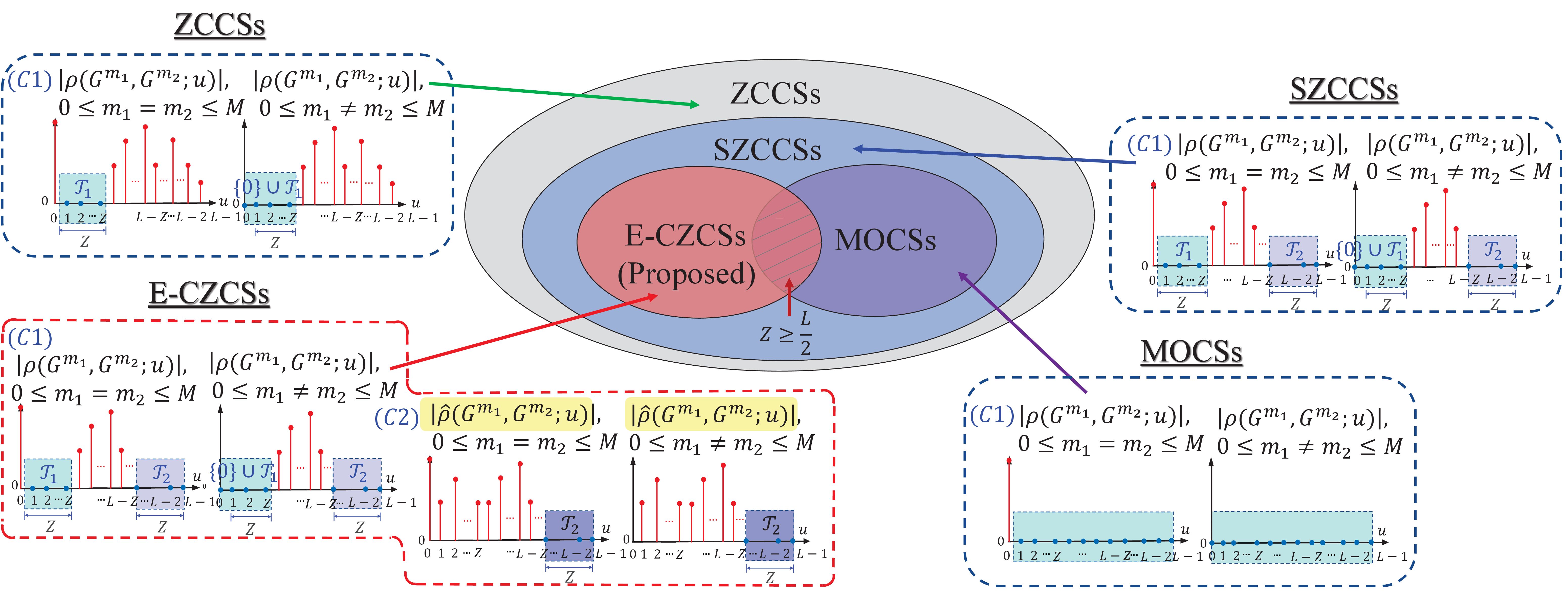}
\caption{Relationship between E-CZCSs and the related sequence sets.\label{E-CZCS_relationship_fig}}
\end{center}
\end{figure*}

Fig.~\ref{E-CZCS_relationship_fig} illustrates the relationship between E-CZCSs and the related sequence sets, which include ZCCSs, SZCCSs, MOCSs, and CCCs.

Next, we discuss the relationship among the ZCZ width $Z$, the set size $M$, and the number of sequences $N$.

\begin{thm}\label{GCZCS_bound}
For an $(M,N,L,Z)$-E-CZCS $\mathcal{G}=\{{G}^{0},{G}^{1},\ldots,{G}^{M-1}\}$, the upper bound on ZCZ width is given by
\begin{equation}
\begin{aligned}
Z\leq \frac{NL}{M}-1.
\end{aligned}\label{GCZCS_bound1}
\end{equation}
For the binary E-CZCS, we have
\begin{equation}
\begin{aligned}
Z\leq \frac{NL}{2M}.
\end{aligned}\label{GCZCS_bound2}
\end{equation}
\end{thm}

\begin{IEEEproof}Let $G^m=\{\bm{g}_{0}^{m},\bm{g}_{1}^{m},$ $\ldots,\bm{g}_{N-1}^{m}\}$ for $m=0,1,\ldots,M-1$ and also let 
\begin{equation}
\begin{aligned}
    &\bm{d}_{0}= \bm{g}_{0}^{0}\|\bm{g}_{1}^{0}\|\cdots\|\bm{g}_{N-1}^{0},\\
    &\bm{d}_{1}= \bm{g}_{0}^{1}\|\bm{g}_{1}^{1}\|\cdots\|\bm{g}_{N-1}^{1},\\
    &{\hspace*{3em}}\vdots\\
    &\bm{d}_{M-1}= \bm{g}_{0}^{M-1}\|\bm{g}_{1}^{M-1}\|\cdots\|\bm{g}_{N-1}^{M-1}.
\end{aligned}\label{Construction_ZCZ1}
\end{equation}
For $m=0,1,\ldots,M-1$, we have
\begin{equation*}
\begin{aligned}
&\phi(\bm{d}_{m};u)=\rho(\bm{g}_{0}^{m};u)+\rho(\bm{g}_{1}^{m};u)+\ldots+\rho(\bm{g}_{N-1}^{m};u)\\
&+\rho^{*}(\bm{g}_{0}^{m},\bm{g}_{1}^{m};L-u)+\rho^{*}(\bm{g}_{1}^{m},\bm{g}_{2}^{m};L-u)+\ldots\\&+\rho^{*}(\bm{g}_{N-1}^{m},\bm{g}_{0}^{m};L-u)\\
&=\sum_{n=0}^{N-1}\rho(\bm{g}_{n}^{m};u)+\sum_{n=0}^{N-1}\rho^{*}(\bm{g}_{n}^{m},\bm{g}_{(n+1)_{\Mod{N}}}^{m};L-u)\\
&=\rho(G^{m},G^{m};u)+\hat{\rho}^{*}(G^{m},G^{m};L-u), \hspace*{1em}\text{for \,}1\leq u< L,
\end{aligned}\label{ZCZ_proof1}
\end{equation*}
and
\begin{equation*}
\begin{aligned}
&\phi(\bm{d}_{m};L)\\&=\rho^{*}(\bm{g}_{0}^{m},\bm{g}_{1}^{m};0)+\rho^{*}(\bm{g}_{1}^{m},\bm{g}_{2}^{m};0)+\ldots+\rho^{*}(\bm{g}_{N-1}^{m},\bm{g}_{0}^{m};0)\\
&=\sum_{n=0}^{N-1}\rho^{*}(\bm{g}_{n}^{m},\bm{g}_{(n+1)_{\Mod{N}}}^{m};0)=\hat{\rho}^{*}(G^{m},G^{m};0).
\end{aligned}\label{ZCZ_proof2}
\end{equation*}
Therefore,
\begin{equation}
\begin{aligned}
\phi(\bm{d}_{m};u)=\begin{cases}
                     \rho(G^{m},G^{m};u)+\hat{\rho}^{*}(G^{m},G^{m};L-u), \\\hspace*{4em} \text{for \, } 1\leq u< L, \quad 0 \leq m \leq M-1; \\
                     \hat{\rho}^{*}(G^{m},G^{m};0), \\\hspace*{4em} \text{for \, } u=L, \quad   0 \leq m \leq M-1.
                   \end{cases}
\end{aligned}\label{ZCZ_proof3}
\end{equation}
Next, for two distinct integers $m_1$,$m_2$ with $0\leq m_1,m_2 \leq M-1$, we have
\begin{equation*}
\begin{aligned}
&\phi(\bm{d}_{m_1},\bm{d}_{m_2};u)=\rho(\bm{g}_{0}^{m_1},\bm{g}_{0}^{m_2};u)+\rho(\bm{g}_{1}^{m_1},\bm{g}_{1}^{m_2};u)+\ldots\\&+\rho(\bm{g}_{N-1}^{m_1},\bm{g}_{N-1}^{m_2};u)+\rho^{*}(\bm{g}_{0}^{m_2},\bm{g}_{1}^{m_1};L-u)\\&+\rho^{*}(\bm{g}_{1}^{m_2},\bm{g}_{2}^{m_1};L-u)+\ldots+\rho^{*}(\bm{g}_{N-1}^{m_2},\bm{g}_{0}^{m_1};L-u)\\
&=\sum_{n=0}^{N-1}\rho(\bm{g}_{n}^{m_1},\bm{g}_{n}^{m_2};u)+\sum_{n=0}^{N-1}\rho^{*}(\bm{g}_{n}^{m_2},\bm{g}_{(n+1)_{\Mod{N}}}^{m_1};L-u)\\
&=\rho(G^{m_1},G^{m_2};u)+\hat{\rho}^{*}(G^{m_2},G^{m_1};L-u), \text{\,for \,}1\leq u< L,
\end{aligned}\label{ZCZ_proof4}
\end{equation*}
\begin{equation*}
\begin{aligned}
&\phi(\bm{d}_{m_1},\bm{d}_{m_2};0)\\
&=\hspace{-0.1em}\rho(\bm{g}_{0}^{m_1},\bm{g}_{0}^{m_2};0)\hspace{-0.1em}+\hspace{-0.1em}\rho(\bm{g}_{1}^{m_1},\bm{g}_{1}^{m_2};0)\hspace{-0.2em}+\ldots\hspace{-0.2em}+\rho(\bm{g}_{N-1}^{m_1},\bm{g}_{N-1}^{m_2};0)\\
&=\sum_{n=0}^{N-1}\rho(\bm{g}_{n}^{m_1},\bm{g}_{n}^{m_2};0)=\rho(G^{m_1},G^{m_2};0),
\end{aligned}\label{ZCZ_proof5}
\end{equation*}
and
\begin{equation*}
\begin{aligned}
&\phi(\bm{d}_{m_1},\bm{d}_{m_2};L)\\
&=\rho^{*}(\bm{g}_{0}^{m_2},\bm{g}_{1}^{m_1};0)+\rho^{*}(\bm{g}_{1}^{m_2},\bm{g}_{2}^{m_1};0)+\ldots\\
&+\rho^{*}(\bm{g}_{N-1}^{m_2},\bm{g}_{0}^{m_1};0)\\
&=\sum_{n=0}^{N-1}\rho^{*}(\bm{g}_{n}^{m_2},\bm{g}_{(n+1)_{\Mod{N}}}^{m_1};0)=\hat{\rho}^{*}(G^{m_2},G^{m_1};0).
\end{aligned}\label{ZCZ_proof6}
\end{equation*}
Thus, we have
\begin{equation}
\begin{aligned}
\phi(\bm{d}_{m_1},\bm{d}_{m_2};u)\hspace{-0.25em}=\hspace{-0.25em}\begin{cases}
                     \rho(G^{m_1},G^{m_2};0),  \quad \text{for \, } u=0, \\ \hspace*{6em} 0 \leq m_1 \neq m_2 \leq M-1; \\
                     \rho(G^{m_1},G^{m_2};u)+\hat{\rho}^{*}(G^{m_2},G^{m_1};L-u), \\
                     \hspace*{8em} \text{for \, } 1\leq u< L, \\\hspace*{5em} \quad 0 \leq m_1 \neq m_2 \leq M-1;\\
                     \hat{\rho}^{*}(G^{m_2},G^{m_1};0),\\ \quad \text{for \, } u=L, \,\, 0 \leq m_1 \neq m_2 \leq M-1.
                   \end{cases}
\end{aligned}\label{ZCZ_proof7}
\end{equation}
From (\ref{GCZCS_defin}), (\ref{ZCZ_proof3}), and (\ref{ZCZ_proof7}), we have
\begin{equation}
\begin{aligned}
\phi(\bm{d}_{m_1},\bm{d}_{m_2};u)=
\begin{cases}
     0,  & \text{for \, }1\leq|u|\leq Z \text{ \, and \, } \\ &0 \leq m_1=m_2 \leq M-1;\\
     0,  & \text{for \, }|u|\leq Z \text{ \, and \, } \\ &0 \leq m_1\neq m_2 \leq M-1
 \end{cases}
 \label{ZCZ_proof8}
\end{aligned}
\end{equation}
since $\mathcal{G}$ is an $(M,N,L,Z)$-E-CZCS. Therefore, the $M$ sequences $\bm{d}_{0}$, $\bm{d}_{1}$, \ldots , $\bm{d}_{M-1}$ form an $(M,NL,Z)$-ZCZ sequence set. According to {\it Lemma \ref{eq:ZCZ_bound}}, the $(M,NL,Z)$-ZCZ sequence set satisfies that $Z \leq (NL)/M-1$ and $Z \leq (NL)/(2M)$ for binary sequence sets. Therefore, we complete the proof.
\end{IEEEproof}

\begin{defin}
A $q$-ary $(M,N,L,Z)$-E-CZCS is called {\it optimal} if $Z=(NL)/M-1$ for $q>2$ or $Z=(NL)/(2M)$ for $q=2$.
\end{defin}

\subsection{E-CZCSs based on ZCCSs}
We first present a construction of GZCZSs based on ZCCSs.

\begin{thm}\label{GCZCS_basic_construction}
Given an $(M,N,L,Z+1)$-ZCCS $\mathcal{S}=\{S^m|0\leq m \leq M-1\}$ where each constitute set $S^m=\{\bm{s}_{n}^{m}|0\leq n \leq N-1\}$. Then, $\mathcal{G}=\{G^m|0\leq m \leq M-1\}$ is an $(M,N,2L,Z)$-E-CZCS by letting
\begin{align*}
    G^m=\{
    &\bm{g}_{0}^m= \bm{s}_{0}^{m}\|\bm{s}_{1}^{m},\\
    &\bm{g}_{1}^m= \bm{s}_{2}^{m}\|\bm{s}_{3}^{m},\\
    &{\hspace*{3em}}\vdots\\
    &\bm{g}_{N/2-1}^m= \bm{s}_{N-2}^{m}\|\bm{s}_{N-1}^{m},\\
    &\bm{g}_{N/2}^m{\hspace*{1em}}= \bm{s}_{0}^{m}\|(-\bm{s}_{1}^{m}),\\
    &\bm{g}_{N/2+1}^m= \bm{s}_{2}^{m}\|(-\bm{s}_{3}^{m}),\\
    &{\hspace*{3em}}\vdots\\
    &\bm{g}_{N-1}^m{\hspace*{1em}}= \bm{s}_{N-2}^{m}\|(-\bm{s}_{N-1}^{m})\}
\end{align*}
for $m=0,1,\ldots,M-1$. Furthermore, if $\mathcal{S}$ is an $(M,N,L)$-MOCS, then $\mathcal{G}$ is an $(M,N,2L,L)$-E-CZCS.
\end{thm}

\begin{IEEEproof} We consider two cases to show that (C1) and (C2) in (\ref{GCZCS_defin}) are satisfied, respectively.

{\it Case 1:} Let $\mathcal{T}_1=\{1,2,\ldots,Z\}$. For $|u|\in \mathcal{T}_1 \cup \{0\}$, we have
\begin{equation*}
\begin{aligned}
&\rho(G^{m_{1}},G^{m_{2}};u)\\&=\sum_{n=0}^{N-1}\rho(\bm g_{n}^{m_1},\bm g_{n}^{m_2}; u)=2\left(\sum_{n=0}^{N-1}\rho(\bm{s}_{n}^{m_1},\bm{s}_{n}^{m_2};u)\right)\\
&+\sum_{n=0}^{\frac{N}{2}-1}\rho^{*}(\bm{s}_{2n}^{m_2},\bm{s}_{2n+1}^{m_1};L-u)-\sum_{n=0}^{\frac{N}{2}-1}\rho^{*}(\bm{s}_{2n}^{m_2},\bm{s}_{2n+1}^{m_1};L-u)
\end{aligned}
\end{equation*}
\begin{equation}\label{thm_proof_1}
\begin{aligned}
=2\left(\sum_{n=0}^{N-1}\rho(\bm{s}_{n}^{m_1},\bm{s}_{n}^{m_2};u)\right)=
 \begin{cases}
     2NL, \hspace{1em} u=0, m_1=m_2; \\
     0, \hspace{1em}  0<|u|\leq Z, m_1=m_2;\\
     0, \hspace{1em}  |u|\leq Z, m_1\neq m_2
 \end{cases}
\end{aligned}
\end{equation}
since $\mathcal{S}=\{S^m|0\leq m \leq M-1\}$ is an $(M,N,L,Z+1)\text{-ZCCS}$. Let $\mathcal{T}_2=\{2L-Z,2L-Z+1,\ldots,2L-1\}$. For $|u|\in \mathcal{T}_2$, we have
\begin{equation*}
\begin{aligned}
&\rho(G^{m_{1}},G^{m_{2}};u)=\sum_{n=0}^{N-1}\rho(\bm g_{n}^{m_1},\bm g_{n}^{m_2}; u)\\&=\sum_{n=0}^{\frac{N}{2}-1}\rho(\bm s_{2n+1}^{m_1},\bm s_{2n}^{m_2}; u-L)-\sum_{n=0}^{\frac{N}{2}-1}\rho(\bm s_{2n+1}^{m_1},\bm s_{2n}^{m_2}; u-L)
\end{aligned}
\end{equation*}
\begin{equation}\label{thm_proof_2}
\begin{aligned}
\hspace*{-4.5em}=0, \hspace*{2em}\text{for any\,}m_1,m_2\in\{0,1,\ldots,M-1\}.
\end{aligned}
\end{equation}
According to (\ref{thm_proof_1}) and (\ref{thm_proof_2}), we obtain that (C1) in (\ref{GCZCS_defin}) holds.

{\it Case 2:} For $|u|\in \mathcal{T}_2$, we have
\begin{equation}
\begin{aligned}
&\hat{\rho}(G^{m_{1}},G^{m_{2}};u)=\sum_{n=0}^{N-1}\rho(\bm g_{n}^{m_1},\bm g_{(n+1)_{\Mod N}}^{m_2};u)\\
&=\sum_{n=0}^{\frac{N}{2}-1}\rho(\bm s_{2n+1}^{m_1},\bm s_{(2n+2)_{\Mod N}}^{m_2};u-L)\\
&-\sum_{n=0}^{\frac{N}{2}-1}\rho(\bm s_{2n+1}^{m_1},\bm s_{(2n+2)_{\Mod N}}^{m_2};u-L)\\
&=0, \hspace*{2em}\text{for any\,\,}m_1,m_2\in\{0,1,\ldots,M-1\}
\end{aligned}
\end{equation}
which means  (C2) in (\ref{GCZCS_defin}) holds. Therefore, $\mathcal{G}$ is an $(M,N,2L,Z)$-E-CZCS.

Moreover, if $\mathcal{S}$ is an $(M,N,L)$-MOCS, we substitute $Z$ by $L$ in {\it Case 1} and {\it Case 2} and hence have $\mathcal{T}_1\cup\mathcal{T}_2=\{1,2,\ldots,2L-1\}$. Therefore, $\mathcal{G}$ is an $(M,N,2L,L)$-E-CZCS.
\end{IEEEproof}

\begin{remark}
To obtain optimal binary E-CZCS, we consider a binary $(M,M,L)$-MOCS $\mathcal{S}$, i.e., $(M,L)$-CCC, in {\it Theorem \ref{GCZCS_basic_construction}}. Then the ZCZ width of the constructed \mbox{$(M,M,2L,L)$-E-CZCS} satisfies the equality given in (\ref{GCZCS_bound2}). Therefore, an optimal binary $(M,M,2L,L)$-E-CZCS can be obtained.
\end{remark}

According to {\it Theorem \ref{GCZCS_basic_construction}}, E-CZCSs can be constructed based on the MOCSs, CCCs, and ZCCSs. Since MOCSs, CCCs, and ZCCSs with various lengths can be obtained from \cite{Wu_21_TIT, Tao_22, Wu_21, Shen_21}, the lengths of the constructed E-CZCSs from {\it Theorem~\ref{GCZCS_basic_construction}} are flexible.

\begin{eg}\label{GCZCS_24_eg}
Let us consider a $(2,2,12,10)$-ZCCS $\mathcal{S}=\{{S}^{0},{S}^{1}\}$ as shown in Table~\ref{ZCCS_12}. A $(2,2,24,9)$-E-CZCS $\mathcal{G}=\{G^0, G^1\}$ can be constructed based on $\mathcal{S}$ according to {\it Theorem~\ref{GCZCS_basic_construction}}. This $(2,2,24,9)$-E-CZCS is listed in Table~\ref{GCZCS_example_basic_2}. We list the correlation sums $\rho(G^{0},G^{0};u)$ and $\hat{\rho}(G^{0},G^{1};u)$ as follows:
\begin{align*}
  \left|\rho(G^{0},G^{0};u)\right|_{\mu=0\sim 23}=(48,\,&{\color{blue}0,0,0,0,0,0,0,0,0},8,\\&0,0,0,0,{\color{blue}0,0,0,0,0,0,0,0,0});
\end{align*}
\begin{align*}
  \left|\hat{\rho}(G^{0},G^{1};u)\right|_{\mu=0\sim 23}=(0,4,&0,4,0,4,16,4,16,4,0,\\&4,0,0,0,{\color{blue}0,0,0,0,0,0,0,0,0}).
\end{align*}
\end{eg}

\begin{table}[!t]
\begin{scriptsize}
\centering
\extrarowheight=3pt
\caption{Binary $(2,2,12,10)$-ZCCS $\mathcal{S}=\{{S}^{0},{S}^{1}\}$\label{ZCCS_12}}
\begin{center}
\begin{tabular}{|l|l|}
\hline
${S}^{0}$ & $\begin{Bmatrix}
\bm s_{0}^{0}&= (++++--+-+-++),\\
\bm s_{1}^{0}&= (-++--+++++-+)
\end{Bmatrix}$ \\ \hline

${S}^{1}$  &$\begin{Bmatrix}
\bm s_{0}^{1}&= (+-+++++--++-),\\
\bm s_{1}^{1}&= (--+-+-++----)
\end{Bmatrix}$  \\ \hline
\end{tabular}
\end{center}
\end{scriptsize}
 \end{table}
 
\begin{table}[!t]
\begin{scriptsize}
\centering
\extrarowheight=3pt
\caption{Binary $(2,2,24,9)$-E-CZCS in Example~\ref{GCZCS_24_eg}\label{GCZCS_example_basic_2}}
\begin{center}
\begin{tabular}{|cc|}
\hline
\multicolumn{2}{|c|}{$(2,2,24,9)$-E-CZCS $\mathcal{G}=\{{G}^{0},{G}^{1}\}$}     \\ \hline
\multicolumn{1}{|c|}{${G}^{0}$} & $\begin{Bmatrix}
\bm g_{0}^{0}=&\hspace*{-2em}(++++--+-+-++\\&-++--+++++-+),\\
\bm g_{1}^{0}=&\hspace*{-2em}(++++--+-+-++\\&+--++-----+-)
\end{Bmatrix}$ \\ \hline
\multicolumn{1}{|c|}{${G}^{1}$} & $\begin{Bmatrix}
\bm g_{0}^{1}=&\hspace*{-2em}(+-+++++--++-\\&--+-+-++----),\\
\bm g_{1}^{1}=&\hspace*{-2em}(+-+++++--++-\\&++-+-+--++++)
\end{Bmatrix}$ \\ \hline
\end{tabular}
\end{center}
\end{scriptsize}
\end{table}

\subsection{E-CZCSs based on Generalized Boolean functions}
In this subsection, we will present a direct construction of \mbox{E-CZCSs} based on generalized Boolean functions. The proposed construction can generate E-CZCSs with various set sizes and large ZCZ widths including optimal binary E-CZCS with the maximum ZCZ width.

\begin{thm}\label{GCZCS_Boolean}
For nonnegative integers $m,v,k$ with $v \leq k$, we let nonempty sets $U_1,U_2,\ldots,U_k$ be a partition of $\{1,2,\ldots,m\}$. Also let $m_{\alpha}$ be the order of $U_{\alpha}$ and $\pi_{\alpha}$ be a bijection from $\{1,2,\ldots,m_{\alpha}\}$ to $U_{\alpha}$ for $\alpha =1,2,\ldots,k$. The generalized Boolean function $f$ is given as
\begin{equation}\label{eq:GBF_GCZCS}
f= \frac{q}{2}\sum_{\alpha=1}^{k}\sum_{\beta=1}^{m_\alpha-1}x_{\pi_\alpha(\beta)}x_{\pi_\alpha(\beta+1)}+\sum_{i=1}^{m}\eta_{i} x_{i}+\eta_{0}
\end{equation}
where ${\eta}_{i} \in \mathbb{Z}_q$ for $i=0,1,\ldots, m$. If $v<k$, we set $\pi_{v+\gamma}(1)=m-\gamma +1$ for $\gamma=1,2,\ldots,k-v$. For $p=0,1,\ldots,2^{k}-1$, we let $G^{p}=\{\zeta_{q}({\bm g}^{p}_0),\zeta_{q}({\bm g}^{p}_1),\ldots,\zeta_{q}({\bm g}^{p}_{2^{v}-1})\}$ where
\begin{equation}\label{eq:GBF_GCZCS_1}
{\bm g}^{p}_{n}={\bm f}+\frac{q}{2}\left(\sum_{\alpha=1}^{v}n_{v-\alpha +1}{\bm x}_{\pi_{\alpha}(1)}+\sum_{\alpha=1}^{k}p_{\alpha}\bm{x}_{\pi_{\alpha}(m_{\alpha})} \right)
\end{equation}
for $n=0,1,\ldots,2^{v}-1$; $(n_1,n_2,\ldots,n_v)$ and $(p_1,p_2,\ldots,p_k)$ are binary representations of $n$ and $p$, respectively. Then the set $\mathcal{G}=\{G^0,G^1,\ldots,G^{2^{k}-1}\}$ is a \mbox{$(2^k,2^v,2^m,2^{\pi_{1}(1)-1})$-E-CZCS}.
\end{thm}

\begin{IEEEproof}
The proof is given in Appendix.
\end{IEEEproof}

\begin{remark}
To obtain the largest ZCZ width, we set $\pi_{1}(1)=m-k+v$ in {\it Theorem} \ref{GCZCS_Boolean} to construct the $(2^k,2^v,2^m,2^{m-k+v-1})$-E-CZCS which can achieve the upper bound on the ZCZ width in (\ref{GCZCS_bound2}). Therefore, optimal binary $(2^k,2^v,2^m,2^{m-k+v-1})$-E-CZCSs can be obtained.
\end{remark}

\begin{eg}\label{GCZCS_32_eg_GBF}
Let us consider $q=2$, $m=5$, $k=2$, and $v=1$. We let a partition of $\{1,2,3,4,5\}$ by $U_1=\{1,2,4\}$ and $U_2=\{3,5\}$ with $m_1=3$ and $m_2=2$, respectively. We also let bijections $\pi_1=(4,1,2)$ and $\pi_2=(5,3)$. Then, the generalized Boolean function $f$ in (\ref{eq:GBF_GCZCS}) can be written as $f = x_{4}x_{1}+x_{1}x_{2}+x_{5}x_{3}$ by setting $\eta_i=0$ for all $i$. Following {\it Theorem \ref{GCZCS_Boolean}}, an optimal binary $(4,2,32,8)$-E-CZCS can be constructed as $\mathcal{G}=\{G^p=\{\zeta_{2}(\bm{g}^{p}_{0}),\zeta_{2}(\bm{g}^{p}_{1})\}:p\in\{0,1,2,3\}\}$ where $\bm{g}^{p}_{n}=\bm{f}+n_1\bm{x}_4+p_1\bm{x}_2+p_2\bm{x}_3$. We list the constituent sequence sets $G^0$, $G^1$, $G^2$, and $G^3$ in Table~\ref{GCZCS_example_GBF}. The correlation sums $\rho(G^{0},G^{2};u)$ and $\hat{\rho}(G^{0},G^{2};u)$ are given as follows:
\begin{align*}
  \left|\rho(G^{0},G^{2};u)\right|_{\mu=0\sim 31}&=({\color{blue}0,0,0,0,0,0,0,0,0},0,0,0,16,0,0,\\&\hspace*{-1em}0,32,0,0,0,16,0,0,0,{\color{blue}0,0,0,0,0,0,0,0});
\end{align*}
\begin{align*}
  \left|\hat{\rho}(G^{0},G^{2};u)\right|_{\mu=0\sim 31}&=(0,0,0,0,0,0,0,0,0,4,0,12,0,12,\\&\hspace*{-2.5em}0,20,0,12,0,4,0,4,0,4,{\color{blue}0,0,0,0,0,0,0,0}).
\end{align*}
\end{eg}


\begin{table*}[!t]
\begin{scriptsize}
\centering
\extrarowheight=3pt
\caption{Optimal Binary $(4,2,32,8)$-E-CZCS in Example \ref{GCZCS_32_eg_GBF} \label{GCZCS_example_GBF}}
\begin{center}
\begin{tabular}{|c||c|c|}
\hline
\multirow{4}{*}{\begin{tabular}[c]{@{}c@{}}$(4,2,32,8)$-E-CZCS\\ $\mathcal{G}=\{{G}^{0},{G}^{1},{G}^{2},{G}^{3}\}$\end{tabular}} & ${G}^{0}$  & $\begin{Bmatrix}
\bm g_{0}^{0}&= (++-+++-++---+-----+-++-+-++++---),\\
\bm g_{1}^{0}&= (++-+++-+-+++-+++--+-++-++----+++),
\end{Bmatrix}$  \\ \cline{2-3}
                   & ${G}^{1}$  & $\begin{Bmatrix}
\bm g_{0}^{1}&= (+++-+++-+-+++-++---++++--+--+-++),\\
\bm g_{1}^{1}&= (+++-+++--+---+-----++++-+-++-+--)
\end{Bmatrix}$ \\ \cline{2-3}
                   & ${G}^{2}$ & $\begin{Bmatrix}
\bm g_{0}^{2}&= (++-+--+-+----+++--+---+--+++-+++),\\
\bm g_{1}^{2}&= (++-+--+--++++-----+---+-+---+---)
\end{Bmatrix}$  \\ \cline{2-3}
                   & ${G}^{3}$ & $\begin{Bmatrix}
\bm g_{0}^{3}&= (+++----++-++-+-----+---+-+---+--),\\
\bm g_{1}^{3}&= (+++----+-+--+-++---+---++-+++-++)
\end{Bmatrix}$ \\ \hline
\end{tabular}
\end{center}
\end{scriptsize}
 \end{table*}

\section{Proposed Training Framework for Broadband GSM Systems}\label{sec:Proposed Training Framework}

\subsection{Training Design}\label{sec:Training Design for Broadband GSM System}

In this subsection, we formulate the system model and the requirements for training design in the GSM system.

\begin{figure}[!t]
\centering
\begin{center}
\extrarowheight=3pt
\includegraphics[width=85mm]{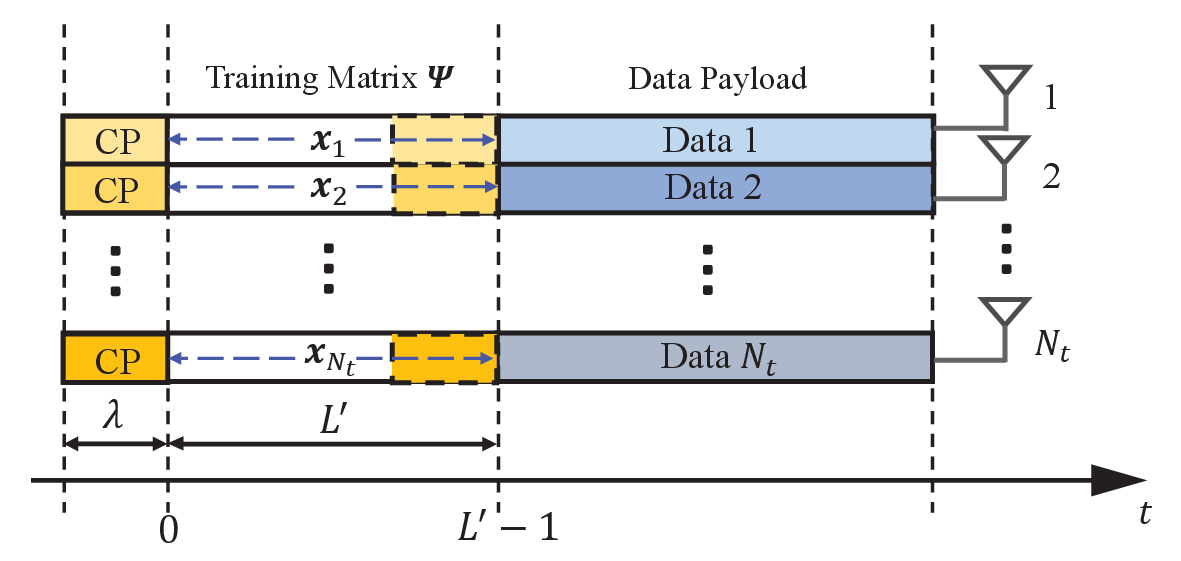}
\caption{A generic training-based SC-MIMO transmission structure.\label{P3}}
\end{center}
\end{figure}

Consider a generic training-based multiple-antenna transmission structure depicted in Fig.~\ref{P3}. Prior to data payload transmission, the training sequences $\bm x_1, \bm x_2, \ldots, \bm x_{N_t}$ transmitted from the $N_t$ TAs are used to estimate the channel state information. The cyclic prefix (CP) is inserted before each training sequence for ISI suppression in dispersive channels. Let $\bm \Psi$ denote the training matrix given by
\begin{equation*}
\begin{aligned}
\bm \Psi={\begin{bmatrix}
   \bm x_1 \\
   \bm x_2 \\
   \vdots \\
   \bm x_{N_t}
 \end{bmatrix}}
 = \left[ {\begin{array}{llll}
  x_{1,0} & x_{1,1} & \ldots & x_{1,L^{\prime}-1} \\
  x_{2,0} & x_{2,1} & \ldots & x_{2,L^{\prime}-1} \\
  \vdots & \vdots & \ddots & \vdots \\
  x_{{N_t},0} & x_{{N_t},1} & \ldots & x_{{N_t},L^{\prime}-1}
\end{array}}\right]_{N_t\times L^{\prime}}\label{CZCSframe}
\end{aligned}
\end{equation*}
where $\bm x_p = (x_{p,0},x_{p,1},\ldots,x_{p,L^{\prime}-1})$ stands for the training sequence conveyed over the $p$-th TA for $p=1,2,\ldots,N_t$. Note that all the training sequences with identical energy $E=\sum_{t=0}^{L^{\prime}-1}|x_{p,t}|^{2}$. In addition, we consider a quasi-static frequency-selective channel with the delay spread $\lambda$. Assume that the channel impulse response from the $p$-th TA to the receiver is $\bm{h}_{p}=[h_{p,0},h_{p,1},\ldots,h_{p,\lambda}]$. To formulate the model in matrix form, we let
\begin{equation}
\begin{aligned}
\bm{X}=[\bm{X}_1,\bm{X}_2,\ldots,\bm{X}_{N_t}]_{L^{\prime}\times N_t(\lambda+1)}
\end{aligned}
\end{equation}
where
\begin{equation*}
\begin{aligned}
\bm{X}_p
 = \left[ {\begin{array}{llll}
  x_{p,0} & x_{p,L^{\prime}-1} & \cdots &  x_{p,L^{\prime}-\lambda}  \\
  x_{p,1} & x_{p,0} & \cdots &  x_{p,L^{\prime}-\lambda+1}  \\
  \vdots & \vdots & \ddots & \vdots \\
  x_{p,L^{\prime}-1} & x_{p,L^{\prime}-2} & \cdots &  x_{p,L^{\prime}-\lambda-1}
\end{array}}\right]_{L^{\prime}\times (\lambda+1)}
\end{aligned}
\end{equation*}
for $p=1,2,\ldots,N_t$. Then, the $L^{\prime} \times 1$ complex received signal vector at a RA can be expressed as
\begin{equation}
\begin{aligned}
\bm{y}=\bm{X}\bm{h}+\bm{w}
\end{aligned}
\end{equation}
where $\bm{h}=[\bm{h}_1, \bm{h}_2, \ldots, \bm{h}_{N_t}]^{T}$ stands for the channel matrix and $\bm w = [w_{0},w_{1},\ldots,w_{L^{\prime}-1}]^T$ stands for the complex additive white Gaussian noise (AWGN) with zero mean and variance $\sigma^2$. By using the LS channel estimator \cite{CZCP-1st, MSE_bound}, the normalized mean square error can be derived as
\begin{align*}
\text{MSE}=\frac{\sigma^2}{N_t\lambda+N_t}\text{Tr}\left(\left(\bm X^H\bm X\right)^{-1}\right).
\end{align*}
Therefore, the minimum MSE can be achieved as ${\sigma^2}/{E}$ \cite{CZCP-1st} if and only if
\begin{equation}
\begin{aligned}
\phi(\bm x_{i}, \bm x_{j}; u)=
\begin{cases}
E,  & \mbox{if }i=j, u=0; \\
0, & \mbox{if }i\neq j, 0\leq u\leq \lambda,\\
& \mbox{or }i= j, 1\leq u\leq \lambda.
\end{cases}\label{optimalcondition}
\end{aligned}
\end{equation}

\begin{remark}
Since the training-based multiple-antenna transmission incorporates the GSM transmission scheme in \mbox{Fig. \ref{GSM_fig}} as a special case with a particular focus on the training matrix design, (\ref{optimalcondition}) is referred to as the optimal condition for GSM training sequences under the LS channel estimator. Furthermore, it should be noted that the training matrix $\bm \Psi$ needs to be sparse since every GSM system only activates a few antennas at each time-slot.
\end{remark}

Hence, the following design criterion provides the optimal channel estimation conditions for the GSM system.

{\it Design criterion:} A training matrix $\bm \Psi$ for the GSM system can achieve the optimal channel estimation over the frequency-selective channel with delay spread $\lambda$, if it satisfies the following two conditions.

(1) Each column of the training matrix $\bm \Psi$ has exactly $N_a$ non-zero entries since $N_a$ TAs are activated over each time-slot in the GSM system.

(2) The training matrix $\bm \Psi$ needs to meet the condition in (\ref{optimalcondition}). 

\subsection{Proposed GSM training matrix} \label{sec:Proposed GSM training matrix}

Based on the design criteria outlined in Subsection \ref{sec:Training Design for Broadband GSM System}, we generate the training matrix employing the proposed E-CZCSs for the broadband GSM system.

For positive integers $N_t$ and $N_a$, we let $V=\lceil\frac{N_t}{N_{a}}\rceil$ where $N_{t}$ is the number of transmit antennas and $N_{\text{a}}$ is the number of RF chains. Let $\bm \Psi_1, \bm \Psi_2,\ldots, \bm \Psi_V$ be the training blocks as follows:
\begin{equation}\label{GSM_matrix_new}
\begin{aligned}
\bm \Psi_1=\begin{bmatrix}
\bm x_1 \\
\bm x_2 \\
\vdots \\
\bm x_{N_a}
\end{bmatrix},
\bm \Psi_2=\begin{bmatrix}
\bm x_{N_a+1} \\
\bm x_{N_a+2} \\
\vdots \\
\bm x_{2N_a}
\end{bmatrix}
\end{aligned},\ldots,
\bm \Psi_V=\begin{bmatrix}
\bm x_{(V-1)N_a+1} \\
\bm x_{(V-1)N_a+2} \\
\vdots \\
\bm x_{VN_a}
\end{bmatrix}.
\end{equation}

Choosing an $(M,N,L,Z)$-E-CZCS $\mathcal{G}$ with the condition $M \geq N_a$, we let $\mathcal{X}_{0}, \mathcal{X}_{1},\ldots, \mathcal{X}_{N-1}$ be the training sub-blocks of size $N_{a}\times VL$ as follows:

\begin{align*}
\mathcal{X}_{0}=\begin{bmatrix}
\bm g_{0}^{0} & \bm 0 &\ldots &\bm 0\\
\bm g_{0}^{1} & \bm 0 &\ldots &\bm 0\\
\vdots & \vdots &\ddots &\vdots\\
\bm g_{0}^{N_{\text{a}}-1} & \bm 0 &\ldots &\bm 0
\end{bmatrix},
\mathcal{X}_{1}=\begin{bmatrix}
\bm g_{1}^{0} & \bm 0 &\ldots &\bm 0\\
\bm g_{1}^{1} & \bm 0 &\ldots &\bm 0\\
\vdots & \vdots &\ddots &\vdots\\
\bm g_{1}^{N_{\text{a}}-1} & \bm 0 &\ldots &\bm 0
\end{bmatrix},
\end{align*}
\begin{align*}
\ldots,
\mathcal{X}_{N-1}=\begin{bmatrix}
\bm g_{N-1}^{0} & \bm 0 &\ldots &\bm 0\\
\bm g_{N-1}^{1} & \bm 0 &\ldots &\bm 0\\
\vdots & \vdots &\ddots &\vdots\\
\bm g_{N-1}^{N_{\text{a}}-1} & \bm 0 &\ldots &\bm 0
\end{bmatrix}
\end{align*}
where $\{\bm{g}_{0}^{0}, \bm{g}_{1}^{0}, \ldots, \bm{g}_{N-1}^{0}\}$, $\{\bm{g}_{0}^{1}, \bm{g}_{1}^{1}, \ldots, \bm{g}_{N-1}^{1}\}$, $\ldots$, $\{\bm{g}_{0}^{N_a-1}, \bm{g}_{1}^{N_a-1}, \ldots, \bm{g}_{N-1}^{N_a-1}\}$ are $N_a$ constituent sequence sets from the $(M,N,L,Z)$-E-CZCS $\mathcal{G}$ and $\bm 0$ represents all-zero vector of length $L$. Then, a $V{N_a}\times NVL$ GSM training matrix $(N_t,N_a,V,N,L)$-$\bm \Psi$ is provided as
\begin{equation}\label{GSM_matrix}
\begin{aligned}
\bm \Psi=\left[ {\begin{array}{c}
\bm \Psi_1 \\
\bm \Psi_2 \\
\vdots \\
\bm \Psi_V
\end{array}}\right]
=\setlength{\arraycolsep}{2pt}\left[ {\begin{array}{cccc}
\mathcal{X}_{0} & \mathcal{X}_{1}& \ldots & \mathcal{X}_{N-1}\\
\mathcal{X}_{0}^{(L)} & \mathcal{X}_{1}^{(L)}& \ldots & \mathcal{X}_{N-1}^{(L)}\\
\vdots & \vdots & \ddots & \vdots\\
\mathcal{X}_{0}^{\left(\left(V-1\right)L\right)} & \mathcal{X}_{1}^{\left(\left(V-1\right)L\right)}& \ldots & \mathcal{X}_{N-1}^{\left(\left(V-1\right)L\right)}\\
\end{array}}\right].
\end{aligned}
\end{equation}
It is noted that in the scenario where $N_t < VN_a$, the first $N_t$ rows of $\bm \Psi$ are selected as training sequences for $N_t$ transmit antennas.

\begin{eg}\label{two_frameworks_eg}
Consider a GSM system equipped with $N_t=4$ TAs and $N_a=2$ RF chains. We have $V=\lceil\frac{N_t}{N_a}\rceil=2$. The \mbox{$(4,2,2,2,L)$-$\bm \Psi$} GSM training matrix based on a \mbox{$(2,2,L,Z)$-E-CZCS} is shown in Fig.~\ref{GCZCS_training_ex}-\subref{GCZCS_4TA_2RF_ex}. If we consider another scenario for the GSM system with $N_t=8$ TAs and $N_a=3$ RF chains (i.e., $V=\lceil\frac{N_t}{N_a}\rceil=3$), the \mbox{$(8,3,3,2,L)$-$\bm \Psi$} GSM training matrix based on a \mbox{$(4,2,L,Z)$-E-CZCS} can be expressed as shown in \mbox{Fig.~\ref{GCZCS_training_ex}-\subref{GCZCS_8TA_4RF_ex}}.
\begin{figure*}[!t]
 \begin{subfigure}{.5\textwidth}
 \parbox[][5cm][c]{\linewidth}{
      \centering
         \includegraphics[width=5.5cm]{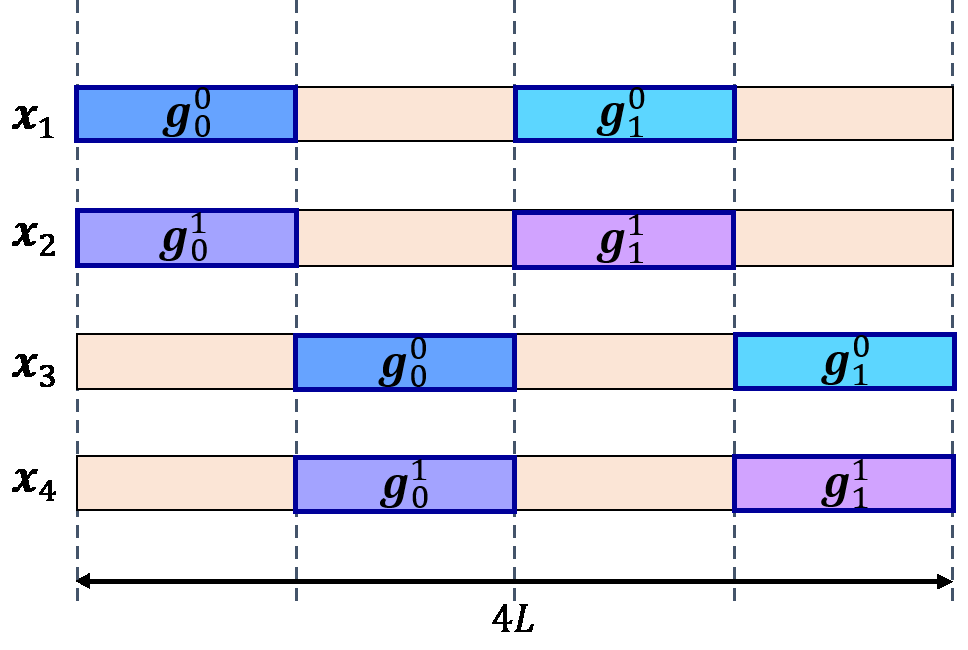}}
         \caption{$(4,2,2,2,L)$-$\bm \Psi$ for $4$ TAs.\label{GCZCS_4TA_2RF_ex}}
 \end{subfigure}
 \begin{subfigure}{.5\textwidth}
         \centering
         \includegraphics[width=7cm]{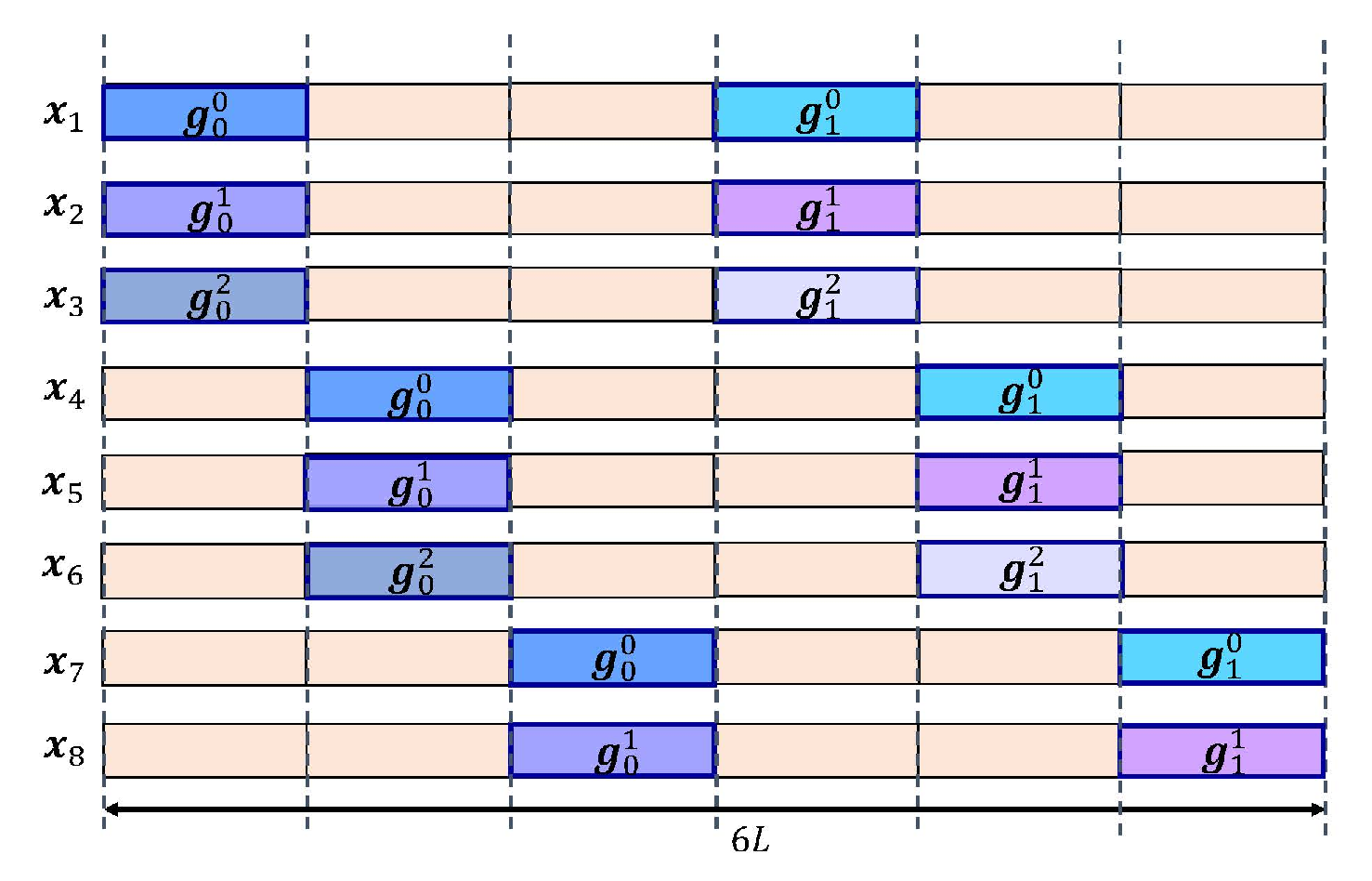}
         \caption{$(8,3,3,2,L)$-$\bm \Psi$ for $8$ TAs.\label{GCZCS_8TA_4RF_ex}}
 \end{subfigure}
\caption{The GSM training matrices based on E-CZCSs in the Example \ref{two_frameworks_eg}.}\label{GCZCS_training_ex}
\end{figure*}
\end{eg}

We then demonstrate that the training matrix $\bm \Psi$ employing the proposed E-CZCSs satisfies the design criteria mentioned in Subsection~\ref{sec:Training Design for Broadband GSM System}. Firstly, there are $N_a$ non-zero entries and $(V-1)N_a$ zeros in each column of the training matrix $\bm \Psi$ as shown in (\ref{GSM_matrix}). Secondly, we prove that the training sequences $\bm x_1, \bm x_2, \ldots, \bm x_{N_t}$, i.e., the first $N_t$ rows of $\bm \Psi$, meet the condition in (\ref{optimalcondition}) if $Z \geq \lambda$. Here, we consider an $(M,N,L,Z)$-E-CZCS $\mathcal{G}=\{G^0,G^1,\ldots,G^{M-1}\}$ with $M \geq N_a$ and $Z \geq \lambda$ where $\lambda$ is the delay spread. The number of transmit antennas $N_{t}$ is divided into $V$ antennas groups $B_1, B_2, \ldots, B_V$ where each $B_v$ consists of $N_a$ antennas where $B_v=\{1+(v-1)N_a,2+(v-1)N_a,\ldots, vN_a\}$ for $v=1,2,\ldots,V$. We consider three cases below to show that the training matrix $\bm \Psi$ satisfies the condition in (\ref{optimalcondition}).

{\it Case 1:}
Since the sets $G^0=\{\bm{g}_{0}^{0}, \bm{g}_{1}^{0}, \ldots, \bm{g}_{N-1}^{0}\}, G^1=\{\bm{g}_{0}^{1}, \bm{g}_{1}^{1}, \ldots, \bm{g}_{N-1}^{1}\}, \ldots, $ and $G^{N_{a}-1}=\{\bm{g}_{0}^{N_a-1}, \bm{g}_{1}^{N_a-1}, \ldots, \bm{g}_{N-1}^{N_a-1}\}$ in $\mathcal{G}$ satisfy the condition (C1) in (\ref{GCZCS_defin}), we have
\begin{equation*}
\begin{aligned}
&\phi\left(\bm{x}_{k},\bm{x}_{l};u\right)=\sum_{j=0}^{N-1}\rho\left(\bm{g}_{j}^{m_{k}}, \bm{g}_{j}^{m_{l}};u\right)\\&=
\begin{cases}
\sum_{j=0}^{N-1}\rho\left(\bm{g}_{j}^{m_{l}}, \bm{g}_{j}^{m_{l}};u\right),\,   \text{for \,}k=l, 0\leq u \leq Z; \\
\sum_{j=0}^{N-1}\rho\left(\bm{g}_{j}^{m_{k}}, \bm{g}_{j}^{m_{l}};u\right),\,    \text{for \,}k\neq l, 0\leq u \leq Z;
\end{cases}\\
&=
\begin{cases}
\rho\left(G^{m_{l}}, G^{m_{l}};u\right),\,  \text{for \,}k=l, 0\leq u \leq Z;  \\
\rho\left(G^{m_{k}}, G^{m_{l}};u\right),\,    \text{for \,}k\neq l, 0\leq u \leq Z;
\end{cases}\\
&=
\begin{cases}
NL,\,\text{for \,}k=l, u=0;\\
0,\,\text{for \,}k=l, 1\leq u \leq Z;\\
0,\,\text{for \,}k\neq l, 0\leq u \leq Z
\end{cases}
\end{aligned}
\end{equation*}
for any $k,l\in B_{v}$ and $v=1,2,\ldots,V$ where $m_{k}=(k-1)_{\Mod {N_a}}$ and $m_{l}=(l-1)_{\Mod {N_a}}$. Over each training block $\bm \Psi_v$, the ISI at each TA and the IAI between the $k$-th and the $l$-th TAs caused by multipath delay can be eliminated.

{\it Case 2:}
For $ k\in B_{v}$, $l\in B_{v+1}$, and $v=1,2,\ldots,V-1$, we have
\begin{equation*}
\begin{aligned}
\phi\left(\bm{x}_{l},\bm{x}_{k};u\right)&=\sum_{j=0}^{N-1}\rho^{*}\left(\bm{g}_{j}^{m_k}, \bm{g}_{j}^{m_l};L-u\right)\\&=
\begin{cases}
\sum_{j=0}^{N-1}\rho^{*}\left(\bm{g}_{j}^{m_k}, \bm{g}_{j}^{m_k};L-u\right),\, \\ \hspace*{10em}\text{if \,}l=k+N_a; \\
\sum_{j=0}^{N-1}\rho^{*}\left(\bm{g}_{j}^{m_k}, \bm{g}_{j}^{m_l};L-u\right),\,  \\ \hspace*{10em} \text{otherwise};
\end{cases}\\
&=
\begin{cases}
\rho^{*}\left(G^{m_k}, G^{m_k};L-u\right),\,  \text{if \,}l=k+N_a;  \\
\rho^{*}\left(G^{m_k}, G^{m_l};L-u\right),\,    \text{otherwise};
\end{cases}\\
&=0
\end{aligned}
\end{equation*}
where $m_{k}=(k-1)_{\Mod {N_a}}$ and $m_{l}=(l-1)_{\Mod {N_a}}$.
The IAI between the $k$-th TA in $\bm \Psi_v$ and the $l$-th TA in $\bm \Psi_{v+1}$ is eliminated.

{\it Case 3:}
For $ k\in B_{1}$ and $l\in B_{V}$, according to (C2) in (\ref{GCZCS_defin}), we have
\begin{equation*}
\begin{aligned}
\phi\left(\bm{x}_{k},\bm{x}_{l};u\right)&=\sum_{j=0}^{N-1}\rho^{*}\left(\bm{g}_{j}^{m_k}, \bm{g}_{(j+1)_{\Mod {N}}}^{m_l};L-u\right)
\end{aligned}
\end{equation*}
\begin{equation*}
\begin{aligned}
&=
\begin{cases}
\sum_{j=0}^{N-1}\rho^{*}\left(\bm{g}_{j}^{m_k}, \bm{g}_{(j+1)_{\Mod {N}}}^{m_k};L-u\right),\,\\ \hspace*{8em}   \text{if \,}l=k+(B-1)N_a; \\
\sum_{j=0}^{N-1}\rho^{*}\left(\bm{g}_{j}^{m_k}, \bm{g}_{(j+1)_{\Mod {N}}}^{m_l};L-u\right),\,\\ \hspace*{8em}   \text{otherwise};
\end{cases}\\
&=
\begin{cases}
\hat{\rho}^{*}\left(G^{m_k}, G^{m_k};L-u\right),\, \\ \hspace*{8em}  \text{if \,}l=k+(B-1)N_a;\\
\hat{\rho}^{*}\left(G^{m_k}, G^{m_l};L-u\right),\, \\ \hspace*{8em}   \text{otherwise};
\end{cases}\\
&=0
\end{aligned}
\end{equation*}
where $m_{k}=(k-1)_{\Mod {N_a}}$ and $m_{l}=(l-1)_{\Mod {N_a}}$. It means that the IAI between the $k$-th TA in $\bm \Psi_1$ and the $l$-th TA in $\bm \Psi_{V}$ is eliminated. From the above three cases, we can conclude that the training matrix $\bm \Psi$ employing the proposed $(M,N,L,Z)$-E-CZCS $\mathcal{G}$ achieves the condition in (\ref{optimalcondition}) if $Z \geq \lambda$.


\section{Simulations}\label{sec:Numerical Evaluation}

\begin{figure*}[!t]
 \begin{subfigure}{.5\textwidth}
 \parbox[][7cm][c]{\linewidth}{
         \centering
         \includegraphics[width=8cm]{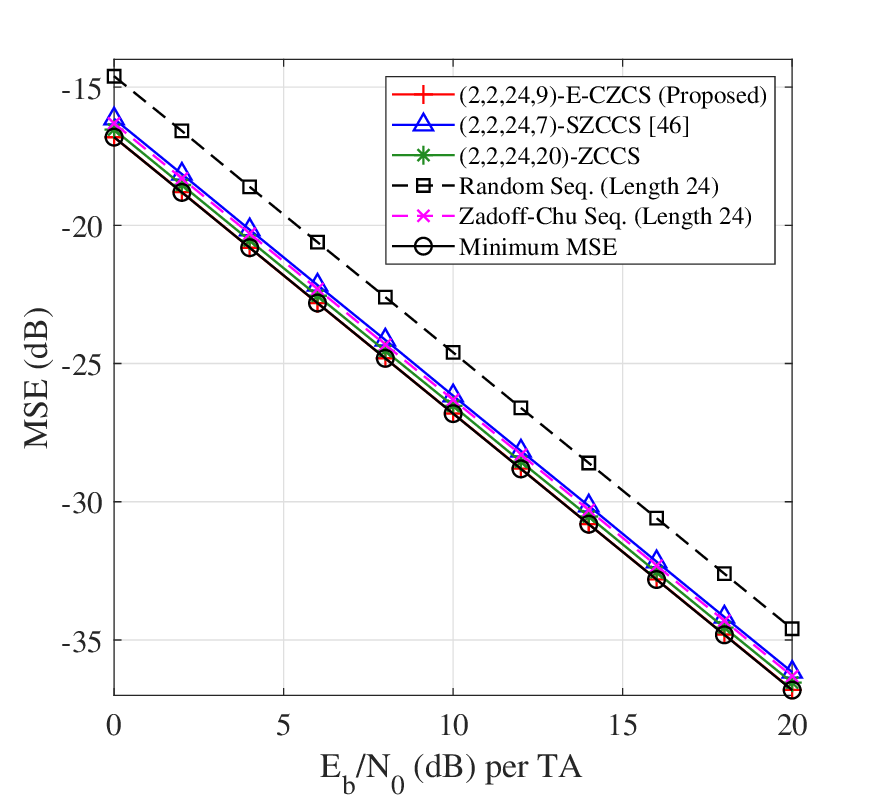}}
         \caption{\label{GCZCS_EbN0_4TA_2RF_24}}
 \end{subfigure}
 \begin{subfigure}{.5\textwidth}
         \centering
         \includegraphics[width=8cm]{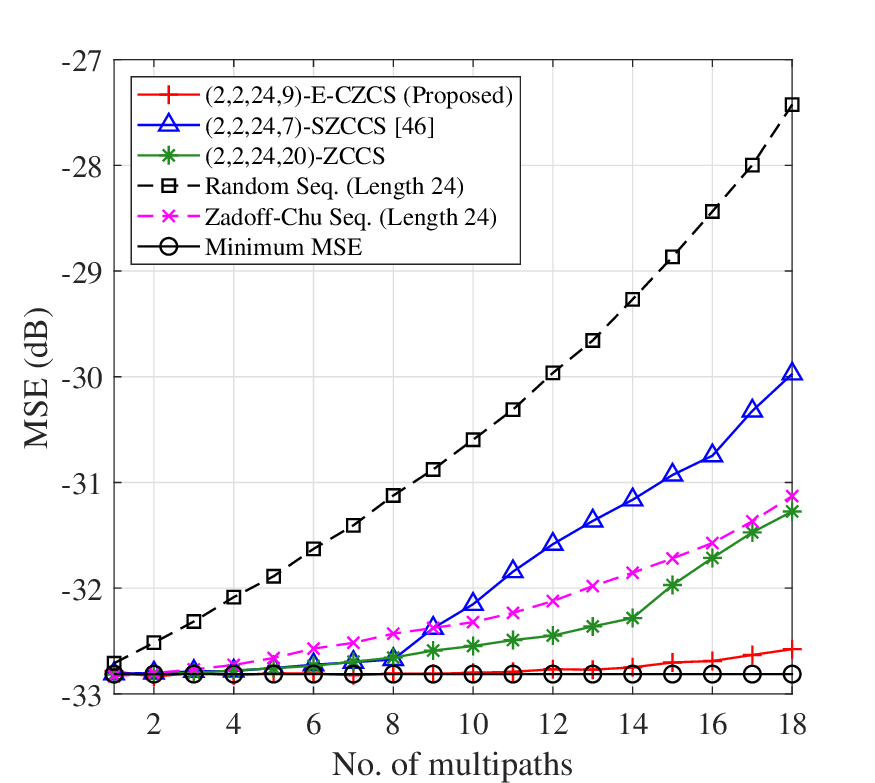}
         \caption{\label{GCZCS_multi_4TA_2RF_24}}
 \end{subfigure}
\caption{MSE comparison of GSM training based on different sequences with $4$ TAs.}\label{GCZCS_4TA_2RF_24}
\end{figure*}

In this section, we examine the channel estimation performance of the proposed E-CZCS-based training for the GSM system over the frequency-selective channel. We consider a $(\lambda+1)\text{-path}$ channel separated by integer symbol durations as $h[t]=\sum_{i=0}^{\lambda}h_i\delta[t-iT]$ where $h_i$'s are complex Gaussian random variables with zero mean and $E\left(|h_i|^2\right)=1/(\lambda+1)$ for all $i$. We evaluate the channel estimation performance of the GSM training matrices based on our proposed E-CZCSs and other classes of sequence sets including the SZCCS, ZCCS,  binary random sequences, and Zadoff-Chu sequences. Our first simulation setup consists of $N_t = 4$ TAs, $N_a = 2$ RF chains, and $N_r = 1$ RA. We employ the binary $(2,2,24,9)\text{-E-CZCS}$ from {\em Example \ref{GCZCS_24_eg}} to generate the $(4,2,2,2,32)$-$\bm \Psi$ as depicted in Fig.~\ref{GCZCS_training_ex}-\subref{GCZCS_4TA_2RF_ex}. For the ZCCS and the SZCCS, the training matrix is given by
\begin{equation}\label{CCCandSZCCS-based}
\begin{aligned}
\left[ {\begin{array}{llll}
\bm{s}_{0}^{0} & \bm{0} & \bm{s}_{1}^{0} & \bm{0} \\
\bm{s}_{0}^{1} & \bm{0} & \bm{s}_{1}^{1} & \bm{0} \\
\bm{0} & \bm{s}_{0}^{0} & \bm{0} & \bm{s}_{1}^{0}  \\
\bm{0} & \bm{s}_{0}^{1} & \bm{0} & \bm{s}_{1}^{1}
\end{array}}\right]_{4\times 128}
\end{aligned}
\end{equation}
where $\{\bm{s}_{0}^{0}, \bm{s}_{1}^{0}\}$ and $\{\bm{s}_{0}^{1}, \bm{s}_{1}^{1}\}$ are the constituent sequence sets of a $(2,2,24,20)$-ZCCS or the $(2,2,24,7)$-SZCCS from \cite{Zhou_23}. For binary random sequences, the elements of $\bm{s}_{0}^{0}$, $\bm{s}_{1}^{0}$, $\bm{s}_{0}^{1}$, and $\bm{s}_{1}^{1}$ in (\ref{CCCandSZCCS-based}) are randomly generated from ``$+1$'' or ``$-1$''. For the training matrix based on Zadoff-Chu sequences, the sequences $\bm{s}_{0}^{0}$, $\bm{s}_{1}^{0}$, $\bm{s}_{0}^{1}$, and $\bm{s}_{1}^{1}$ are assigned by $4$ distinct \mbox{Zadoff-Chu sequences} of length $24$ with low cross-correlations. The MSE performances of the channel estimation based on different training matrices as shown in Fig.~\ref{GCZCS_4TA_2RF_24}. Fig.~\mbox{\ref{GCZCS_4TA_2RF_24}-\subref{GCZCS_EbN0_4TA_2RF_24}} shows the MSE performances of the channel estimation versus the changes of $E_{b}/N_{0}$ with $\lambda=9$. It can be seen that the MSE performance of the proposed GSM training matrix with the $(2,2,24,9)\text{-E-CZCS}$ matches the MSE lower bound. In Fig.~\mbox{\ref{GCZCS_4TA_2RF_24}-\subref{GCZCS_multi_4TA_2RF_24}}, we consider different numbers of multipaths at $E_{b}/N_{0}=16$ dB. When the number of multipaths is less than or equal to $10$, i.e., $\lambda=9$, our proposed E-CZCS-based training matrix can achieve the minimum MSE since the ZCZ width is $9$. For the SZCCS-based GSM, the performance is worse when the number of multipaths is larger than 8. This is because the SZCCS only consider the condition (C1) in (\ref{GCZCS_defin}) and the ZCZ width is only $7$. When the delay spread is larger than $7$, the out-of-zone correlations of the SZCCS degrade the channel estimation performance. For the ZCCS-based training, the condition (C2) in (\ref{GCZCS_defin}) is not met, thus leading nonzero IAI. 

\begin{figure}[!t]
\centering
\includegraphics[width=8cm]{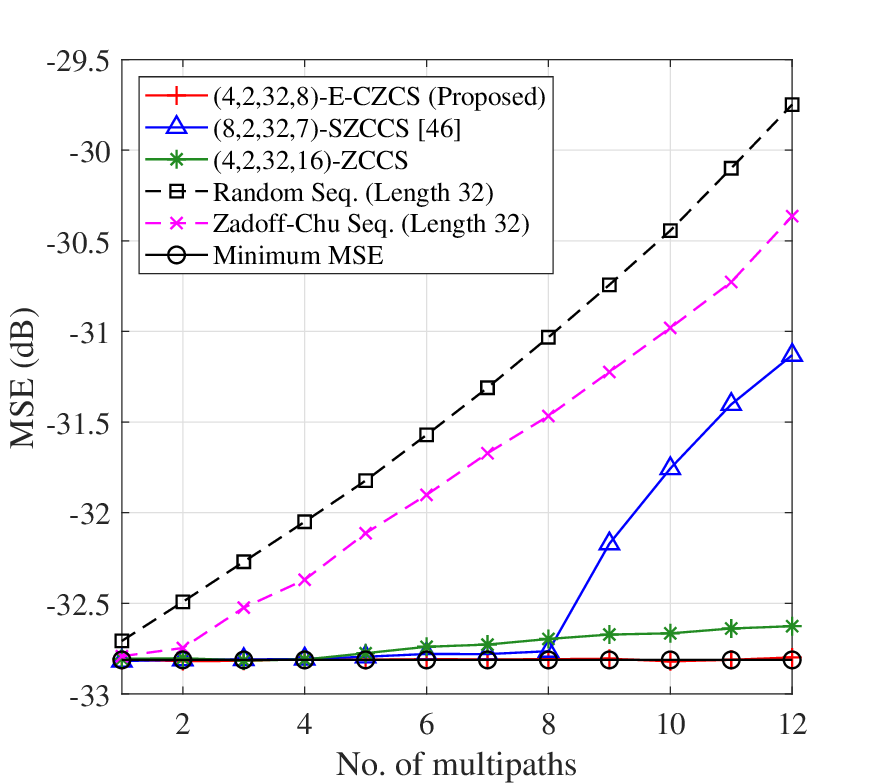}
\caption{MSE comparison of GSM training based on different sequences with $8$ TAs.}\label{GCZCS_multi_8TA_3RF}
\end{figure}

In Fig.~\ref{GCZCS_multi_8TA_3RF}, we consider the GSM system with $N_t = 8$ TAs, $N_a = 3$ RF chains, and $N_r = 1$ RA. We use the GSM training matrix $(8,3,3,2,32)$-$\bm \Psi$ as depicted in Fig.~\mbox{\ref{GCZCS_training_ex}-\subref{GCZCS_8TA_4RF_ex}} based on the optimal binary $(4,2,32,8)$-E-CZCS from {\em Example~\ref{GCZCS_32_eg_GBF}}. For comparison, we take the first three sequence sets of the $(8,2,32,7)$-SZCCS from \cite{Zhou_23} into consideration, as well as the $(4,2,32,16)$-ZCCS, binary random sequences, and \mbox{Zadoff-Chu} sequences. The training matrix is given by
\begin{equation}\label{CCCandSZCCS-based-8TA4RF}
\begin{aligned}
\left[ {\begin{array}{llllll}
\bm{s}_{0}^{0} & \bm{0} & \bm{0} & \bm{s}_{1}^{0} & \bm{0} & \bm{0} \\
\bm{s}_{0}^{1} & \bm{0} & \bm{0} & \bm{s}_{1}^{1} & \bm{0} & \bm{0} \\
\bm{s}_{0}^{2} & \bm{0} & \bm{0} & \bm{s}_{1}^{2} & \bm{0} & \bm{0} \\
\bm{0} & \bm{s}_{0}^{0} & \bm{0} & \bm{0} & \bm{s}_{1}^{0} & \bm{0} \\
\bm{0} & \bm{s}_{0}^{1} & \bm{0} & \bm{0} & \bm{s}_{1}^{1} & \bm{0} \\
\bm{0} & \bm{s}_{0}^{2} & \bm{0} & \bm{0} & \bm{s}_{1}^{2} & \bm{0} \\
\bm{0} & \bm{0} & \bm{s}_{0}^{0} & \bm{0} & \bm{0} &\bm{s}_{1}^{0}\\
\bm{0} & \bm{0} & \bm{s}_{0}^{1} & \bm{0} & \bm{0} &\bm{s}_{1}^{1}
\end{array}}\right]_{8\times 192}
\end{aligned}
\end{equation}
where the component sequences $\bm{s}_{n}^{m}$'s are assigned in a similar manner as in the previous simulation. For example, $\{\bm{s}_{0}^{0},\bm{s}_{1}^{0}\}$, $\{\bm{s}_{0}^{1}, \bm{s}_{1}^{1}\}$, and $\{\bm{s}_{0}^{2}, \bm{s}_{1}^{2}\}$ are the three constituent sets of the $(4,2,32,16)\text{-ZCCS}$ if the training matrix is based on the ZCCS. For Fig.~\ref{GCZCS_multi_8TA_3RF}, the $E_{b}/N_{0}$ is fixed at $16$ dB. We observe that the $(4,2,32,8)$-E-CZCS outperforms others and still performs close to the MSE lower bound even the delay spread is larger than $8$. This is because the out-of-zone correlations of the proposed $(4,2,32,8)$-E-CZCS are small. For the SZCCS based training, the performance degrades significantly when the delay spread is larger than the ZCZ width, e.g., the delay spread $\lambda \geq 8$.


\section{Conclusion}\label{sec:conclusion}

This paper is focused on a novel class of sequence sets called E-CZCSs, each consisting of a collection of CZCSs and with an additional cross-channel aperiodic correlation sum property. We have proposed two systematic constructions of E-CZCSs including one based on ZCCSs, MOCSs, and CCCs ({\it Theorem \ref{GCZCS_basic_construction}}) and the other based on generalized Boolean functions ({\it Theorem \ref{GCZCS_Boolean}}). Both constructions can generate the optimal binary E-CZCSs.

Furthermore, a novel GSM training framework has been proposed based on E-CZCSs. It is shown that the proposed training design can achieve the optimal channel estimation performance in frequency-selective channels, by fully exploiting the correlation properties of E-CZCS. 

Although {\it Theorem \ref{GCZCS_Boolean}} can generate E-CZCSs with various set sizes and ZCZ widths, the lengths are currently limited to powers of two. Therefore, a potential future direction is to construct E-CZCSs with non-power-two sequence lengths.

\appendix[Proof of Theorem \ref{GCZCS_Boolean}]\label{apxC}

Before proving {\it Theorem \ref{GCZCS_Boolean}}, we introduce the following lemma which can be used to prove our main theorem.

\begin{lemma}\cite{Chen08}\label{eq:CCC_k_path_GBF}
For any integer $m$ and $k$ with $0 < k\leq m$, let nonempty sets $U_1,U_2,\ldots ,U_k$ be a partition of $\{1,2,\ldots,m\}$. Also let $\pi_{\alpha}$ be a bijection from $\{1,2,\ldots,m_{\alpha}\}$ to $U_{\alpha}$ where $m_{\alpha}$ is the order of $U_{\alpha}$ for $\alpha =1,2,\ldots,k$. Given an even positive integer $q$ and the generalized Boolean function $f$
\begin{equation}\label{eq:CCC_k_path_GBF_1}
f= \frac{q}{2}\sum_{\alpha=1}^{k}\sum_{\beta=1}^{m_\alpha-1}x_{\pi_\alpha(\beta)}x_{\pi_\alpha(\beta+1)}+\sum_{i=1}^{m}\eta_{i} x_{i}+\eta_{0}
\end{equation}
where ${\eta}_{i}$'s $\in \mathbb{Z}_q$. For $0\leq \kappa, \nu\leq 2^k-1$, the set $C^\nu=\{\bm{c}_{0}^{\nu},\bm{c}_{1}^{\nu},\ldots,\bm{c}_{2^{k}-1}^{\nu}\}$ can be constructed as follows:
\begin{equation}\label{eq:CCC_k_path_GBF_2}
\bm{c}_{\kappa}^{\nu}= \bm{f}+\frac{q}{2}\sum_{\alpha=1}^{k}\kappa_{\alpha}\bm{x}_{\pi_{\alpha}(1)} +\frac{q}{2}\sum_{\alpha=1}^{k}\nu_{\alpha}\bm{x}_{\pi_{\alpha}(m_{\alpha})}
\end{equation}
where $(\kappa_1,\kappa_2,\ldots,\kappa_k)$ and $(\nu_1,\nu_2,\ldots,\nu_k)$ are binary representations of $\kappa$ and $\nu$, respectively. Then, $C^0, C^1, \ldots, C^{2^{k}-1}$ form a $(2^k,2^m)$-CCC.
\end{lemma}

\begin{IEEEproof}[Proof of Theorem \ref{GCZCS_Boolean}] We consider three parts to illustrate that $\mathcal{G}$ satisfies (C1) and (C2) in (\ref{GCZCS_defin}) where $\mathcal{T}_1=\{1,2,\ldots,2^{\pi_1(1)-1}\}$ and $\mathcal{T}_2=\{2^m-2^{\pi_1(1)-1},2^m-2^{\pi_1(1)-1}+1,\ldots,2^{m}-1\}$. Let ${\bm g}^{p}_{n}=({g}^{p}_{n,0},{g}^{p}_{n,1},\ldots,{g}^{p}_{n,L-1})$ for $p=0,1,\ldots,2^{v}-1$ and $n=0,1,\ldots,2^{k}-1$.

In the first part, we have to demonstrate that
\begin{equation*}
\begin{aligned}
&\hspace*{2em}\rho(G^p,G^p;u)=\sum_{n=0}^{2^v-1} \rho\left({\zeta_q(\bm g}^{p}_{n}),\zeta_q({\bm g}^{p}_{n});u\right)\\
&=\sum_{n=0}^{2^v-1}\sum_{i=0}^{2^{m}-1-u}\xi_{q}^{{g}^{p}_{n,i+u}-{g}^{p}_{n,i}}=\sum_{i=0}^{2^{m}-1-u}\sum_{n=0}^{2^v-1}\xi_{q}^{{g}^{p}_{n,i+u}-{g}^{p}_{n,i}}=0,
\end{aligned}
\end{equation*}
for $|u| \in \mathcal{T}_1 \cup \mathcal{T}_2$. If $v=k$, the sequences $\bm{g}_{n}^{p}$ in (\ref{eq:GBF_GCZCS_1}) can be rewritten as
\begin{equation}\label{eq:GCZCS_set}
{\bm g}^{p}_{n}={\bm f}+\frac{q}{2}\left(\sum_{\alpha=1}^{k}n_{k-\alpha +1}{\bm x}_{\pi_{\alpha}(1)}+\sum_{\alpha=1}^{k}p_{\alpha}\bm{x}_{\pi_{\alpha}(m_{\alpha})} \right)
\end{equation}
implying $G^{p} \in \mathcal{G}$ is a GCS as given in {\it Lemma \ref{eq:CCC_k_path_GBF}}. Hence, we have $\rho(G^p,G^p;u)=0$, $|u|\in\{1,2,\ldots,2^{m}-1\}$.

If $v<k$, we consider two cases to show that $\rho(G^p,G^p;u)=\sum_{n=0}^{2^v-1} \rho\left(\zeta_q({\bm g}^{p}_{n});u\right)=0$ when $|u|\in \mathcal{T}_1$ and $|u|\in \mathcal{T}_2$, respectively. For a nonnegative integer $i$ with binary representation $(i_1,i_2,\ldots,i_m)$, we let $j=i+u$ with binary representation $(j_1,j_2,\ldots,j_m)$.

{\it Case 1-A:} We assume $i_{\pi_{1}(1)}\neq j_{\pi_{1}(1)}$ in this case. For any sequence ${\bm g}^{p}_n\in G^{p}$ where $0\leq p\leq 2^{k}-1$ and $0\leq n\leq 2^{v}-1$, there exists a sequence ${\bm g}^{p}_s={\bm g}^{p}_n+(q/2){\bm x}_{\pi_{1}(1)}\in G^{p}$ such that
\begin{equation*}
\begin{aligned}
{g}^{p}_{n,j}-{g}^{p}_{n,i}-{g}^{p}_{s,j}+{g}^{p}_{s,i}&=\frac{q}{2}(i_{\pi_{1}(1)}-j_{\pi_{1}(1)})\equiv \frac{q}{2} \pmod q.
\end{aligned}
\end{equation*}
Since $i_{\pi_{1}(1)}\neq$ $j_{\pi_{1}(1)}$, we can obtain
\begin{equation}
\xi_{q}^{{g}^{p}_{n,j}-{g}^{p}_{n,i}}/\xi_{q}^{{g}^{p}_{s,j}-{g}^{p}_{s,i}}=\xi_{q}^{\frac{q}{2}(i_{\pi_{1}(1)}-j_{\pi_{1}(1)})}=e^{\frac{j2\pi}{q}\frac{q}{2}}=-1
\end{equation}
implying $\xi^{{g}^{p}_{n,j}-{g}^{p}_{n,i}}+\xi^{{g}^{p}_{s,j}-{g}^{p}_{s,i}}=0$. Therefore, we have $\sum_{n=0}^{2^v-1}\xi_{q}^{{g}^{p}_{n,j}-{g}^{p}_{n,i}}=0$.

{\it Case 1-B:} In this case we have $i_{\pi_{1}(1)}=$ $j_{\pi_{1}(1)}$ and we can deduce that $i_{\pi_{v+\gamma}(1)}=j_{\pi_{v+\gamma}(1)}$ for $\gamma=1,2,\ldots,k-v$. Suppose not, let $\alpha^{\prime}$ be the smallest integer satisfying  $i_{\pi_{v+\alpha^{\prime}}(1)}\neq$ $j_{\pi_{v+\alpha^{\prime}}(1)}$. Therefore, $i_{m}=j_{m}, i_{m-1}=j_{m-1}, \ldots, i_{m-\alpha^{\prime}+2}=j_{m-\alpha^{\prime}+2}$. Then,
\begin{equation*}
\begin{aligned}
u=j-i&=2^{m-\alpha^{\prime}}+\sum_{s=1,~s\neq \pi_{1}(1)}^{m-\alpha^{\prime}}(j_s-i_s)2^{s-1}\\
&\geq 2^{m-\alpha^{\prime}}-\sum_{s=1}^{m-\alpha^{\prime}}2^{s-1}+2^{\pi_{1}(1)-1}=2^{\pi_{1}(1)-1}+1
\end{aligned}
\end{equation*}
which contradicts the assumption that $u\leq 2^{\pi_{1}(1)-1}$. So we have $i_{\pi_{v+1}(1)}=j_{\pi_{v+1}(1)}, i_{\pi_{v+2}(1)}=j_{\pi_{v+2}(1)}, \ldots, i_{\pi_{k}(1)}=j_{\pi_{k}(1)}$ here. Then we consider two subcases below.

{\it Case 1-B (i):} We assume $i_{\pi_{\alpha}(1)}\neq$ $j_{\pi_{\alpha}(1)}$ for some $\alpha=2,3,\ldots,v$. For any sequence ${\bm g}^{p}_n\in G^{p}$, there exists another sequence ${\bm g}^{p}_s={\bm g}^{p}_n+(q/2){\bm x}_{\pi_{\alpha}(1)}\in G^{p}$ such that $\xi_{q}^{{g}^{p}_{n,j}-{g}^{p}_{n,i}}+\xi_{q}^{{g}^{p}_{s,j}-{g}^{p}_{s,i}}=0$.

{\it Case 1-B (ii):} Following the above case, we have $i_{\pi_{\alpha}(1)}=j_{\pi_{\alpha}(1)}$ for all $\alpha=1,2,\ldots,k$. We assume $i_{\pi_{\alpha}(\beta)}=j_{\pi_{\alpha}(\beta)}$ for $\alpha=1,2\ldots,\hat{\alpha}-1$ with $\hat{\alpha}\leq k$ and $\beta=1,2,\ldots,m_{\alpha}$. Then we suppose that $\hat{\beta}$ is the smallest integer such that $i_{\pi_{\hat{\alpha}}(\hat{\beta})}\neq$ $j_{\pi_{\hat{\alpha}}(\hat{\beta})}$. Let $i'$ and $j'$ be two integers which are distinct from $i$ and $j$, respectively, only in one position $\pi_{\hat{\alpha}}(\hat{\beta}-1)$. That is, ${i'}_{\pi_{\hat{\alpha}}(\hat{\beta}-1)}=1-i_{\pi_{\hat{\alpha}}(\hat{\beta}-1)}$ and ${j'}_{\pi_{\hat{\alpha}}(\hat{\beta}-1)}=1-j_{\pi_{\hat{\alpha}}(\hat{\beta}-1)}$. Hence, we have
\begin{equation*}
\begin{aligned}
&{g}^{p}_{n,i'}-{g}^{p}_{n,i}\\&=\frac{q}{2}\Big({i}_{\pi_{\hat{\alpha}}(\hat{\beta}-2)}{i'}_{\pi_{\hat{\alpha}}(\hat{\beta}-1)}\hspace*{-0.2em}-{i}_{\pi_{\hat{\alpha}}(\hat{\beta}-2)}{i}_{\pi_{\hat{\alpha}}(\hat{\beta}-1)}\hspace*{-0.2em}+{i'}_{\pi_{\hat{\alpha}}(\hat{\beta}-1)}{i}_{\pi_{\hat{\alpha}}(\hat{\beta})}\\
&-{i}_{\pi_{\hat{\alpha}}(\hat{\beta}-1)}{i}_{\pi_{\hat{\alpha}}(\hat{\beta})}\Big)+\eta_{\pi_{\hat{\alpha}}(\hat{\beta}-1)}{i'}_{\pi_{\hat{\alpha}}(\hat{\beta}-1)}+\eta_{\pi_{\hat{\alpha}}(\hat{\beta}-1)}{i}_{\pi_{\hat{\alpha}}(\hat{\beta}-1)}\\
&\equiv\frac{q}{2}\left({i}_{\pi_{\hat{\alpha}}(\hat{\beta}-2)}+{i}_{\pi_{\hat{\alpha}}(\hat{\beta})}\right)+\eta_{\pi_{\hat{\alpha}}(\hat{\beta}-1)}\left(1-2{i}_{\pi_{\hat{\alpha}}(\hat{\beta}-1)}\right)\\& \hspace*{20em} \pmod q.
\end{aligned}
\end{equation*}
Since $i_{\pi_{\hat{\alpha}}(\hat{\beta}-2)}=j_{\pi_{\hat{\alpha}}(\hat{\beta}-2)}$ and $i_{\pi_{\hat{\alpha}}(\hat{\beta}-1)}=j_{\pi_{\hat{\alpha}}(\hat{\beta}-1)}$, we have
\begin{equation}
\begin{aligned}
{g}^{p}_{n,j}-{g}^{p}_{n,i}-{g}^{p}_{n,j'}+{g}^{p}_{n,i'}&\equiv\frac{q}{2}\left({i}_{\pi_{\hat{\alpha}}(\hat{\beta})}-{j}_{\pi_{\hat{\alpha}}(\hat{\beta})}\right)\\&\equiv\frac{q}{2} \pmod q.
\end{aligned}
\end{equation}
Then, we can obtain
\begin{equation*}
\xi_{q}^{{g}^{p}_{n,j}-{g}^{p}_{n,i}}/\xi_{q}^{{g}^{p}_{n,j'}-{g}^{p}_{n,i'}}=\xi_{q}^{\frac{q}{2}({i}_{\pi_{\hat{\alpha}}(\hat{\beta})}-{j}_{\pi_{\hat{\alpha}}(\hat{\beta})})}=e^{\frac{j2\pi}{q}\frac{q}{2}}=-1.
\end{equation*}
Therefore, $\xi_{q}^{{g}^{p}_{n,j}-{g}^{p}_{n,i}}+\xi_{q}^{{g}^{p}_{n,j'}-{g}^{p}_{n,i'}}=0$.

From Case 1-A and Case 1-B, we can conclude that $\rho(G^p,G^p;u)=0, \text{\, for \,} |u|\in \mathcal{T}_1$.

{\it Case 2:} Then, let us consider $|u|\in \mathcal{T}_2$, i.e., $2^{m}-2^{\pi_{1}(1)-1} \leq |u|\leq 2^{m}-1$. In this case, we should have $i_{\pi_{1}(1)}\neq j_{\pi_{1}(1)}$. Suppose not. If $i_{\pi_{1}(1)}= j_{\pi_{1}(1)}$, then we have
\begin{equation*}
\begin{aligned}
u=j-i=\sum_{s=1,s\neq \pi_{1}(1)}^{m}(j_s-i_s)2^{s-1}\leq 2^{m}-2^{\pi_{1}(1)-1}-1
\end{aligned}
\end{equation*}
which contradicts the assumption $2^{m}-2^{\pi_{1}(1)-1} \leq |u|\leq 2^{m}-1$. Hence, we must have $i_{\pi_{1}(1)}\neq j_{\pi_{1}(1)}$ here. Following the similar arguments as given in {\it Case 1-A}, we can also obtain $\xi_{q}^{{g}^{p}_{n,j}-{g}^{p}_{n,i}}+\xi_{q}^{{g}^{p}_{s,j}-{g}^{p}_{s,i}}=0$ where $\bm{g}^{p}_{s}=\bm{g}^{p}_{n}+(q/2)\bm{x}_{\pi_{1}(1)}\in G^p$. Therefore,
\begin{equation*}
\rho(G^p,G^p;u)=\sum_{n=0}^{2^{v}-1}\sum_{i=0}^{2^{m}-1-u}\xi_{q}^{g_{n,j}^{p}-g_{n,i}^{p}}=0, \text{\, for \,} |u|\in \mathcal{T}_2.
\end{equation*}

In the second part, we will demonstrate that any two distinct constituent sets $G^p$ and $G^l$, where $0 \leq p\neq l \leq 2^{k}-1$, have zero cross-correlation sum for $|u| \in \mathcal{T}_1 \cup \mathcal{T}_2$, i.e.,
\begin{equation*}
\begin{aligned}
&\hspace*{2em}\rho(G^p,G^l;u)=\sum_{n=0}^{2^v-1} \rho\left(\zeta_q({\bm g}^{p}_{n}),\zeta_q({\bm g}^{l}_{n});u\right)\\
&=\sum_{n=0}^{2^v-1}\sum_{i=0}^{2^{m}-1-u}\xi_{q}^{{g}^{p}_{n,i+u}-{g}^{l}_{n,i}}=\sum_{i=0}^{2^{m}-1-u}\sum_{n=0}^{2^v-1}\xi_{q}^{{g}^{p}_{n,i+u}-{g}^{l}_{n,i}}=0.
\end{aligned}
\end{equation*}
For $v=k$, similar to the first part, we can obtain that $G^{p}$ and $G^{l}$ are mutually orthogonal GCSs, i.e., $\rho(G^p,G^l;u)=0, \quad |u|\in\{0,1,2,\ldots,2^{m}-1\}$. For $v<k$, by following the similar arguments in the first part, we can obtain that
\begin{equation}
\sum_{n=0}^{2^{v}-1}\rho\left(\zeta_q(\bm{g}_{n}^{p}),\zeta_q(\bm{g}_{n}^{l});u\right)=0, \text{\, for \,} |u|\in \mathcal{T}_1\cup \mathcal{T}_2.
\end{equation}
Now, it only suffices to show that
\begin{equation}
\begin{aligned}
\rho(G^p,G^l;0)
=\sum_{n=0}^{2^v-1}\sum_{i=0}^{2^{m}-1}\xi_{q}^{{g}^{p}_{n,i}-{g}^{l}_{n,i}}=0.
\end{aligned}
\end{equation}
We denote $p_{\alpha}$ and $l_{\alpha}$ as the $\alpha$-th bits of the binary representations of $p$ and $l$, respectively. Also, let $i_{\pi_{\alpha}(m_{\alpha})}$ be the ${\pi_{\alpha}}(m_{\alpha})$-th bit of the binary representation of $i$. According to (\ref{eq:GBF_GCZCS_1}), we have
\begin{equation}\label{uzero}
{g}^{p}_{n,i}-{g}^{l}_{n,i}\equiv \frac{q}{2}\sum_{\alpha=1}^{k}\left(p_{\alpha}-l_{\alpha}\right)i_{\pi_{\alpha}(m_{\alpha})} \pmod q.
\end{equation}
It can be observed that (\ref{uzero}) is the linear combination of the term $i_{\pi_{\alpha}(m_{\alpha})}$. For $i$ ranging from $0$ to $2^{m}-1$, there are $2^{m-1}$ $i$'s such that $\xi_{q}^{{g}^{p}_{n,i}-{g}^{l}_{n,i}}=\xi_{q}^{q/2}=-1$ and $2^{m-1}$ $i$'s such that $\xi_{q}^{{g}^{p}_{n,i}-{g}^{l}_{n,i}}=\xi_{q}^{0}=1$. Therefore, we can obtain $\sum_{i=0}^{2^{m}-1}\xi_{q}^{{g}^{p}_{n,i}-{g}^{l}_{n,i}}=0$.

In the last part, we will prove the condition (C2) in (\ref{GCZCS_defin}) holds for $\mathcal{G}$, i.e.,
\begin{equation}\label{last_part_proof}
\begin{aligned}
&\hat{\rho}(G^p,G^l;u)=\sum_{n=0}^{N-1} \rho\left(\zeta_q({\bm g}^{p}_{n}),\zeta_q({\bm g}^{l}_{(n+1)_{\Mod N}});u\right)\\
&=\sum_{i=0}^{2^{m}-1-u}\sum_{n=0}^{N-1}\xi_{q}^{{g}^{p}_{n,i+u}-{g}^{l}_{{(n+1)_{\Mod N}},i}}=0
\end{aligned}
\end{equation}
where $2^{m}-2^{\pi_{1}(1)-1} \leq |u|\leq 2^{m}-1$ and $N=2^v$. Similarly, we let $j=i+u$ for any integer $i$. From {\it Case 2} in the first part, we deduce that $i_{\pi_{1}(1)}\neq j_{\pi_{1}(1)}$. Let $(n_1,n_2,\ldots,n_v)$ and $(h_1,h_2,\ldots,h_v)$ be the binary representations of $n$ and $h=(n+1)_{\Mod N}$, respectively. We also let $n'$ and $h'$ be the integers that are distinct from $n$ and $h$, respectively, in only one position, i.e., $n^{\prime}_{v}=1-n_v$ and $h^{\prime}_{v}=1-h_v$. We can obtain
\begin{equation*}
\begin{aligned}
g^{p}_{n,j}-g^{p}_{n',j}&=\frac{q}{2}\left(n_{v}j_{\pi_{1}(1)}-(1-n_{v})j_{\pi_{1}(1)}\right)\\
&=-\frac{q}{2}j_{\pi_{1}(1)}+q n_v j_{\pi_{1}(1)}\equiv -\frac{q}{2}j_{\pi_{1}(1)}\pmod q
\end{aligned}
\end{equation*}
and
\begin{equation*}
\begin{aligned}
g^{l}_{h,i}-g^{l}_{h',i}&=\frac{q}{2}\left(h_{v}i_{\pi_{1}(1)}-(1-h_{v})i_{\pi_{1}(1)}\right)\\
&=-\frac{q}{2}i_{\pi_{1}(1)}+q h_v i_{\pi_{1}(1)}\equiv -\frac{q}{2}i_{\pi_{1}(1)}\pmod q.
\end{aligned}
\end{equation*}
Then, we have
\begin{equation*}
\begin{aligned}
g^{p}_{n,j}-g^{l}_{h,i}-g^{p}_{n',j}+g^{l}_{h^{\prime},i}&\equiv\frac{q}{2}(i_{\pi_{1}(1)}-j_{\pi_{1}(1)})\equiv \frac{q}{2}\hspace*{-0.5em} \pmod q
\end{aligned}
\end{equation*}
since $i_{\pi_1(1)}\neq j_{\pi_1(1)}$. Therefore, $\xi_{q}^{g^{p}_{n,j}-g^{l}_{h,i}}+\xi_{q}^{g^{p}_{n',j}-g^{l}_{h^{\prime},i}}=0$ and (\ref{last_part_proof}) is proved. From the above three parts, we can conclude that $\mathcal{G}$ is a $(2^k,2^v,2^m,2^{\pi_1(1)-1})$-E-CZCS.
\end{IEEEproof}

\bibliographystyle{IEEEtran}
\bibliography{IEEEabrv,GCZCS}
\end{document}